\documentclass[a4paper, amsfonts, floatfix, amssymb, amsmath, reprint, showkeys, nofootinbib, oneside, superscriptaddress, preprintnumbers]{revtex4-1}

\pdfoutput=1

\usepackage[english]{babel}
\usepackage[utf8]{inputenc}
\usepackage{slashed}
\usepackage{lipsum} 
\usepackage[force]{feynmp-auto}
\usepackage{graphicx}
\usepackage{dcolumn}
\usepackage{bm}
\usepackage{color}
\usepackage{multirow}
\usepackage{textgreek}
\usepackage{tabularx}
\usepackage{slashed}
\usepackage{simpler-wick}

\usepackage{multirow, makecell}
\newcolumntype{C}{>{\centering\arraybackslash}X}
\usepackage[colorinlistoftodos, color=green!40, prependcaption]{todonotes}
\usepackage{amsthm}
\usepackage{mathtools}
\usepackage{physics}
\usepackage{xcolor}
\usepackage{graphicx}
\usepackage[left=23mm,right=13mm,top=35mm,columnsep=15pt]{geometry} 
\usepackage{adjustbox}
\usepackage{placeins}
\usepackage[T1]{fontenc}
\usepackage{lipsum}
\usepackage{float}
\usepackage{csquotes}
\usetikzlibrary{decorations.pathmorphing}

\usepackage[pdftex, pdftitle={Article}, pdfauthor={Author}]{hyperref} % For hyperlinks in the PDF
\DeclareUnicodeCharacter{2009}{\,} 

\renewcommand{\>}{\rangle}
\newcommand{\<}{\langle}

\newcommand{\cev}[1]{\reflectbox{\ensuremath{\vec{\reflectbox{\ensuremath{#1}}}}}}

\begin{document}
\title{Lattice QCD calculation of the Compton amplitude subtraction function}

\author{K.~U.~Can}
    \affiliation{CSSM, Department of Physics, The University of Adelaide, Adelaide SA 5005, Australia}
\author{A.~Hannaford-Gunn}
    \affiliation{CSSM, Department of Physics, The University of Adelaide, Adelaide SA 5005, Australia}
\author{R.~Horsley}
    \affiliation{School of Physics and Astronomy, University of Edinburgh, Edinburgh EH9 3JZ, UK}
\author{P.~E.~L.~Rakow}
    \affiliation{Theoretical Physics Division, Department of Mathematical Sciences, University of Liverpool, Liverpool L69 3BX, UK}
\author{T.~Schar}
    \affiliation{CSSM, Department of Physics, The University of Adelaide, Adelaide SA 5005, Australia}
\author{G.~Schierholz}
    \affiliation{Deutsches Elektronen-Synchrotron DESY, Notkestr.~85, 22607 Hamburg, Germany}
\author{H.~St\"uben}
    \affiliation{Regionales Rechenzentrum, Universit\"at Hamburg, 20146 Hamburg, Germany}
\author{R.~D.~Young}
    \affiliation{CSSM, Department of Physics, The University of Adelaide, Adelaide SA 5005, Australia}
\author{J.~M.~Zanotti}
    \affiliation{CSSM, Department of Physics, The University of Adelaide, Adelaide SA 5005, Australia}
%    \author{J.~Crawford}
%    \affiliation{CSSM, Department of Physics, The University of Adelaide, Adelaide SA 5005, Australia}

    \collaboration{QCDSF Collaboration}
	\noaffiliation

\date{\today} % Leave empty to omit a date

\begin{abstract}

The Compton amplitude subtraction function is an essential component in work concerning both the proton radius puzzle and the proton-neutron mass difference. However, owing to the difficulty in determining the subtraction function, it remains a key source of uncertainty in these two contexts. Here, we use the Feynman-Hellmann method to determine this subtraction function directly from lattice QCD. Furthermore, we demonstrate how to control dominant discretisation artefacts for this calculation, eliminating a major source of systematic error. This calculation is performed for a range of hard momentum scales, and three different sets of gauge configurations for pion masses about $400\;\text{MeV}$. Our results show good agreement with continuum OPE expectations. As such, this work paves the way for model-independent and precise determinations of the subtraction function over a wide range of kinematics.

\end{abstract}

\keywords{}
\preprint{ADP-24-23/T1262}
\preprint{DESY-24-221}
\preprint{Liverpool LTH 1390}
\maketitle

\section{Introduction}
\label{sec:intro}

The Compton amplitude subtraction function, $S_1(Q^2)$, is a critical component in the theoretical determination of the proton--neutron mass difference, and a necessary quantity for the calculation of the proton charge radius. However, the subtraction function is not directly measurable from experiment, and as such contributes significant theoretical and statistical uncertainties to these quantities. In this context, a first-principles determination of the subtraction function could play a vital role in clarifying our understanding of hadron structure.

The leading electromagnetic contribution to the proton--neutron mass difference is given by the Cottingham sum rule \cite{Cottingham_1963}
\begin{equation}
    \delta m^{\gamma}_N = -\frac{i}{2m_p}\frac{\alpha}{(2\pi)^3} \int d^4 q \, \frac{g_{\mu\nu}T_N^{\mu\nu}(p,q)}{Q^2 -i\epsilon},
    \label{massdiff}
\end{equation}
where $T^N_{\mu\nu}$ is the spin-averaged forward Compton scattering amplitude for a nucleon $N(p)$
\begin{equation}\label{compdef}
    T^N_{\mu \nu}(p,q) = i\int d^4z \  e^{i q\cdot z} \langle N(p)| \mathcal{T} \{ V_\mu(z)V_\nu(0) \} | N(p)\rangle,
\end{equation}
with $V_\mu$ denoting the electromagnetic vector current and $q$ denoting the virtual photon 4-momentum with $Q^2=-q^2$. The subtraction function $S_1(Q^2)$, defined below, encapsulates critical information about $T^N_{\mu\nu}$ at the $p \cdot q = 0$ point. 

An early evaluation of the Cottingham sum rule by Gasser and Leutwyler \cite{Gasser:1974} found the electromagnetic contribution to the mass difference to be $\delta m_N^\gamma = 0.76(30)\;\text{MeV}$. By contrast, a more recent evaluation \cite{Walker-Loud_2012} proposed a significantly greater value, largely due to a different consideration of the Compton subtraction function. As a consequence of these developments, renewed interest was taken in the proton--neutron mass difference \cite{Thomas:2014, Erben:2014, Tomalak:2018, Horsley_2016, Borsanyi_2015}. In light of further clarification \cite{Gasser_2020_main}, focusing particularly on the treatment of the subtraction function, the exact value of $\delta m_N^\gamma$ remains somewhat unclear. Thus a first-principles evaluation of $S_1(Q^2)$ would be rather beneficial.

%By contrast, a more recent evaluation \cite{Walker-Loud_2012} produced a substantially larger value of $m_{\text{QED}}=1.30(03)(47)\;\text{MeV}$, although the outcome was rather limited by knowledge of the subtraction function. Furthermore, this result was subsequently corrected by \cite{Gasser_2020_main} and reduced to a central value of $m_{\text{QED}}=0.63 \; \text{MeV}$. Nevertheless, as a consequence of these developments, renewed interest was taken in the proton-neutron mass difference \cite{Thomas:2014, Erben:2014, Tomalak:2018}. 

Similarly, the Compton amplitude is an input for measurements of the proton charge radius from the muonic-hydrogen Lamb shift, for which recent determinations have a 7$\sigma$ tension with results obtained via electron--proton scattering \cite{Bernauer_2010, Pohl:2010}---the so-called `proton radius puzzle' \cite{Pohl:2013yb, Antognini:2022xoo}. This has led to renewed interest in possible systematic corrections to the muonic-hydrogen measurements, including the hadronic corrections from the two-photon exchange (TPE) \cite{Tomalak:2015, Afanasev:2017, Pachucki_1999}, which are dependent on the forward Compton amplitude and, consequently, the subtraction function. Unfortunately, the Compton amplitude contributes the dominant uncertainty to the hadronic background, in large part due to $S_1(Q^2)$ \cite{Carlson_2011, Paz_2011, Miller:2012}. Therefore a more precise determination of the subtraction function could also help clarify the proton radius puzzle.

In both the context of the proton-neutron mass difference, and the muonic-hydrogen Lamb shift, knowledge of the Compton amplitude over the entire range of virtual photon momenta is essential. The subtraction function can be evaluated in a model independent way in the low-energy region ($Q^2\ll m_p^2$) from effective field theory \cite{Paz_2011}, and in the high-energy region ($Q^2\gg \Lambda_{\text{QCD}}^2$) using the operator product expansion (OPE) \cite{Collins_1979,Hill_2017,Gasser_2020}. In the intermediate $Q^2$ region various different methods have been applied including physically-motivated functional interpolations between the low and high $Q^2$ \cite{Walker-Loud_2012, Erben:2014, Hill_2017}, and Regge model evaluations \cite{Gasser:1974, Gasser_2020, Caprini_2021}. The different methods of evaluating the unknown behaviour of the subtraction function lead to significant differences in both the proton-neutron mass difference and the two-photon exchange hadronic background.

In this context, a first-principles lattice QCD evaluation of the Compton amplitude subtraction function is crucial. Recently a lattice calculation of the subtraction function was performed using four-point correlation functions \cite{Xu_new}, following on from earlier work \cite{Feng:2018qpx, tpelattice4pt}. This calculation was performed over the range $Q^2 \in [0,2] \, \text{GeV}^2$ at the physical pion mass and at an alternative subtraction point \cite{Hagelstein_2021}. The resultant subtraction function was determined from a combination of lattice results and experimental data. 

%Original citation below: \cite{hannafordgunn2021generalised, offfwd2}
In this paper, we determine the Compton amplitude subtraction function over the range $1 \lesssim Q^2 \lesssim 16 \, \text{GeV}^{2}$ at the standard subtraction point using the Feynman-Hellmann technique, which has previously been used to determine the structure function moments for the forward \cite{fwdletter, fwdpaper, fwdpaperscaling} and off-forward \cite{hannafordgunn2021generalised, Hannaford-Gunn:2024aix} Compton amplitudes. Preliminary results of our work on the Feynman-Hellmann subtraction function were first published in Ref.~\cite{interlacingsubtraction}. Unlike Ref.~\cite{Xu_new}, we do not incorporate experimental data into our subtraction function calculation, but rather present pure lattice results. 

The structure of this paper is as follows: In Section \ref{sec:bckgrnd} we discuss in further detail the definition of the subtraction function and its parameterisation in the high-energy region. In Section \ref{sec:feynhell} we discuss the Feynman-Hellmann method used to determine the Compton amplitude, highlighting aspects of this implementation most relevant to the subtraction function, and initial results. Then, in Section \ref{sec:lope}, we discuss our method for correcting leading discretisation artefacts in our calculation. This is done using a lattice operator product expansion (LOPE). Finally, in Section \ref{sec:proton} we remove the leading order discretisation artefacts and, in doing so, present the Compton amplitude subtraction function, and compare it with the continuum OPE expectation.

\section{Background} 
\label{sec:bckgrnd}

\begin{figure}
\centering
\begin{tikzpicture}[thick, scale=1.5]
    % Define the central vertex as a circle
    \node[circle, draw, fill=gray!20, minimum size=1cm] (v) {};

    % Sinusoidal line at 45 degrees (shorter and longer wavelength)
    \draw[decorate, decoration={snake, amplitude=2pt, segment length=10pt}] 
        (v.45) -- ++(0.7,0.7) node[midway, below, yshift=-4pt, xshift=+5pt] {$q$};

    % Sinusoidal line at 135 degrees (shorter and longer wavelength)
    \draw[decorate, decoration={snake, amplitude=2pt, segment length=10pt}] 
        (v.135) -- ++(-0.7,0.7) node[midway, below, yshift=-4pt, xshift=-5pt] {$q$};

    % Arrows 
    \draw[->, thick] (0.3,0.55) -- (0.6,0.85);   % (outwards)
    \draw[->, thick] (-0.6,0.85) -- (-0.3,0.55); % (inwards)

    % Double line at 215 degrees
    \draw[thick] (-0.8,-0.8) -- (-0.23,-0.23) node[midway, above, xshift=-6.5pt] {$p$};
    \draw[thick] (-0.76,-0.84) -- (-0.19,-0.27);

    % Double line at 315 degrees
    \draw[thick] (0.8,-0.8) -- (0.23,-0.23) node[midway, above, xshift=6.5pt] {$p$};
    \draw[thick] (0.76,-0.84) -- (0.19,-0.27);

    % Draw a black triangle
    \fill[black] (-0.4,-0.44) -- (-0.53,-0.71) -- (-0.67,-0.57) -- cycle;

    \fill[black] (0.6,-0.64) -- (0.47,-0.37) -- (0.33,-0.51) -- cycle;

\end{tikzpicture}
\caption{The diagram for proton forward Compton scattering, $\gamma^{*} (q)p(p)  \to \gamma^{*}(q)p(p)$.}
\label{comppic}
\end{figure}
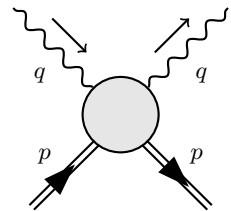

Our starting point is the forward virtual Compton amplitude for a proton, as shown in Eq. \eqref{compdef}. This amplitude describes the process of photon-proton scattering, $\gamma^{*} (q)p(p)  \to \gamma^{*}(q)p(p)$---see Fig.~\ref{comppic}. By performing the standard tensor decomposition, the Compton amplitude can be written in the form \cite{fwdpaper}
\begin{equation}\label{tensdecomp}
    \begin{split}
        &T_{\mu \nu } = \left(-g_{\mu\nu} + \frac{q_\mu q_\nu}{q^2} \right)T_1(\omega,Q^2) 
        \\ & + \frac{1}{p \cdot q}\left(p_\mu - \frac{p \cdot q}{q^2} q_\mu \right)\left(p_\nu - \frac{p \cdot q}{q^2} q_\nu \right)T_2(\omega, Q^2),
    \end{split}
\end{equation}
where $\omega = 2p\cdot q/Q^2$ and $Q^2=-q^2$. The alternative decomposition put forward in Ref.~\cite{Gasser_2020} could provide an opportunity for future work. 

The two Compton amplitude structure functions, $T_{1,2}$, satisfy the following dispersion relations \cite{pasquini_DRs1, pasquini_DRs2}:
\begin{align}
   \begin{split}
        T_1(\omega,Q^2) & = T_1(0,Q^2) + \frac{2\omega^2}{\pi}\int_{1}^{\infty}d\omega' \frac{\text{Im}T_1(\omega',Q^2)}{\omega'(\omega'^2-\omega^2-i\epsilon)},
        \\ T_2(\omega,Q^2) & = \frac{2\omega}{\pi}\int_{1}^{\infty}d\omega' \frac{\text{Im}T_2(\omega',Q^2)}{\omega'^2-\omega^2-i\epsilon}.
        \label{T12_disp}
   \end{split}
\end{align}
Crucially, the imaginary parts of $T_{1,2}$ can be determined from deep-inelastic scattering. On the other hand, $T_1(0,Q^2)$ can not be experimentally accessed from inclusive processes. This is the \emph{subtraction function}:
\begin{equation}
S_1(Q^2) \equiv T_1(0,Q^2) .
    \label{subdef}
\end{equation}
Although the subtraction function is not experimentally accessible, it can be estimated in both the low $Q^2$ and high $Q^2$ regions. 

For $Q^2\ll m_p^2$ the subtraction function has been computed in a variety of formalisms, including effective field theory calculations \cite{Alarcon:2013cba, Caprini:2016wvy, Caprini_2021, Carlson_2011, Erben:2014, Paz_2011, Gasser:1974, Miller:2012, Thomas:2014, Tomalak:2015aoa, Tomalak:2018, Walker-Loud_2012, Walker-Loud:2019}. These calculations have sizeable errors and are not always consistent with one another \cite{Gasser_2020}. 
%FOR LOW Q^2 APPROX: \textcolor{red}{Up to where in $Q^2$ is this valid? Also what about dispersive/model-dependent evaluations?}

In the region $Q^2\gg\Lambda_{\text{QCD}}^2$ the subtraction function can be evaluated model-independently using the operator product expansion (OPE).
Such an expansion is performed in powers of $m_p^2/Q^2$, with the predicted asymptotic behaviour
\begin{equation}
    S_1(Q^2) \sim \frac{m_p^2}{Q^2}, \quad \text{for}\quad Q^2\gg \Lambda_{\text{QCD}}^2.
    \label{opepred}
\end{equation}

Collins \cite{Collins_1979} originally determined the leading-order $m_p^2/Q^2$ contribution from the OPE. Hill and Paz later corrected this calculation, again only evaluating the leading $m_p^2/Q^2$ term. Up to gluon contributions which are suppressed by $\alpha_s$, the OPE result can be written \cite{Hill_2017}
\begin{equation}
S^{\text{OPE}}_1(Q^2) = \frac{4m_p^2}{Q^2}\sum_q \mathcal{Q}_q^2\left ( \< x \>_q - \frac{m_q}{m_p}g^q_S\right),
    \label{hillpaz_s1}
\end{equation}
where $\<x\>_q$ is the second parton distribution function moment (i.e.~the momentum fraction), $m_q$ is the quark mass and $g^q_S$ is the scalar charge---all for a quark of flavour $q$ and charge $\mathcal{Q}_q$. 

Note that the expression in Eq.~\eqref{hillpaz_s1} is only the leading-order in $m_p^2/Q^2$ contribution to the OPE prediction. There are higher order terms in $m_p^2/Q^2$ that may contribute significantly, especially for $Q^2$ on the threshold of the perturbative region.

At low energies, in addition to the OPE contribution, there is also an elastic contribution to the subtraction function \cite{Walker-Loud_2012, Erben:2014, gasserleutwyler, Gasser_2020_main}

\section{Feynman-Hellmann method}
\label{sec:feynhell}

The Feynman-Hellmann (FH) method is a powerful tool that presents an alternative to the direct calculation of three- and four-point functions in lattice QCD. In particular, Feynman-Hellmann has proven particularly important in the case of matrix elements with two current insertions such as the Compton amplitude \cite{fwdletter, fwdpaper, fwdpaperscaling, hannafordgunn2021generalised, Hannaford-Gunn:2024aix}. 

Here we summarise the main aspects of our application of the Feynman-Hellmann method, noting that the full details can be found in Ref.~\cite{fwdpaper}. The starting point is to perturb the action, $\mathcal{S}$, by introducing external fields that couple to the valence quarks of the nucleon. For a particular choice of external field, Eq.~\eqref{action}, the ground-state energy of the nucleon in the external field can be related to the Compton amplitude, and thus the subtraction function, as demonstrated in Ref.~\cite{fwdpaper}. To access the flavour-diagonal piece and the flavour-mixed piece of the Compton amplitude, we perturb the action by
\begin{align}
\begin{split}
    \mathcal{S}(\lambda) &= \mathcal{S} + \lambda_q \int d^3z \,  \left( e^{i\mathbf{q}\cdot\mathbf{z}}+e^{-i\mathbf{q}\cdot\mathbf{z}}\right) V^q_{\mu}(z),\\
    \mathcal{S}(\lambda) &= \mathcal{S} + \sum_{q}\lambda_q \int d^3z \,  \left( e^{i\mathbf{q}\cdot\mathbf{z}}+e^{-i\mathbf{q}\cdot\mathbf{z}}\right) V^q_{\mu}(z)
\end{split}\label{action}
\end{align}
respectively, noting that $\mathbf{q}$ is the 3-momentum carried by the external current, and $V^q_{\mu} = Z_V\bar{\psi}^q\gamma_{\mu}\psi^q$ is the local Euclidean vector current of the quark $q=u,d$ for some choice of Lorentz index $\mu$ and with $Z_V$ denoting the renormalisation constant.

We implement these action modifications on the level of quark propagators
\begin{equation}
    S^q_{\lambda_q}(z_n - z_m) = \big[ M-\lambda_q \mathcal{O} \big]^{-1}_{n,m},
    \label{q_prop}
\end{equation}
where $M$ is the fermion matrix, $n,m$ are lattice sites and $\lambda_q$ is the FH coupling. 

With the quark propagator in Eq.~\eqref{q_prop}, the perturbed proton correlators for different flavour combinations are written generally, up to spin/colour structure, as
\begin{align}
    \begin{split}
        \mathcal{G}(\lambda_u, \lambda_d) \sim \langle S^u_{\lambda_u} S^u_{\lambda_u} S^d_{\lambda_d} \rangle
        \label{pert_prot_corr}
    \end{split}
\end{align}
where, for example, the flavour combination $uu$ is given by $\mathcal{G}(\lambda_u,0) \sim \langle S^u_{\lambda_u} S^u_{\lambda_u} S^d\rangle$, with $S_d$ denoting the unperturbed down-quark propagator. Applying the general principles of Refs.~\cite{fwdpaper, fwdpaperscaling}, we can write the ground state energy of the nucleon in the external field as
\begin{align}
    E(\lambda_u, \lambda_d) &= m_p - \frac{\lambda_u^2}{2m_p}S^{uu}_1(Q^2) - \frac{\lambda_d^2}{2m_p}S^{dd}_1(Q^2) \nonumber\\
    &- \frac{\lambda_u \lambda_d}{2m_p}\big[ S^{ud}_1(Q^2) + S^{du}_1(Q^2) \big] + ...,\label{energy}
\end{align}
noting that the ellipsis denotes terms of higher order in $\lambda$, $S^{ud}_1(Q^2) = S^{du}_1(Q^2)$, and where $S^{qq'}_1(Q^2)$ is the contribution to the subtraction function from 
\begin{equation}\label{compdef_2}
    T^{qq'}_{\mu \nu}(p,q) = i\int d^4z \  e^{i q\cdot z} \langle N(p)| \mathcal{T} \{ V^q_\mu(z)V^{q'}_\nu(0) \} | N(p)\rangle.
\end{equation}

% straight to relation from proceedings (2nd order FH) - energy shift 
% Eq. (27) from proceedings
% Ross' energy equation
% After Eq. (9), second line vanishes when q=q'
% Put into words original 2nd order derivatives for S^qq'

This Feynman-Hellmann relation is only strictly valid when the sink momentum is $\mathbf{p} = \boldsymbol{0}$. Information on the $\mathbf{p} \ne \boldsymbol{0}$ case can be found in Ref.~\cite{fwdpaper}. By definition $q_4=0$ in the FH procedure and hence the probe scale is given by $Q^2 = |\mathbf{q}|^2$. As explained further in the following section, in our calculations we imposed a sink momentum of $\mathbf{p} = \boldsymbol{0}$ which, with $q_4=0$, ensures $\omega=0$. 

The proton subtraction function, up to disconnected contributions, is then constructed via
\begin{equation}
    \begin{split}
        & S_1^p(Q^2) = \sum_{q,q' = u,d}\mathcal{Q}_{q}\mathcal{Q}_{q'}S^{qq'}_1(Q^2) 
         \\ &= \frac{4}{9} S_1^{uu}(Q^2) + \frac{1}{9}S_1^{dd}(Q^2)-\frac{2}{9}\bigg(S_1^{ud}(Q^2) + S_1^{du}(Q^2)\bigg).
         \label{proton_sub}
    \end{split}
\end{equation}

%_______________________________________________________________________
%_______________________________________________________________________

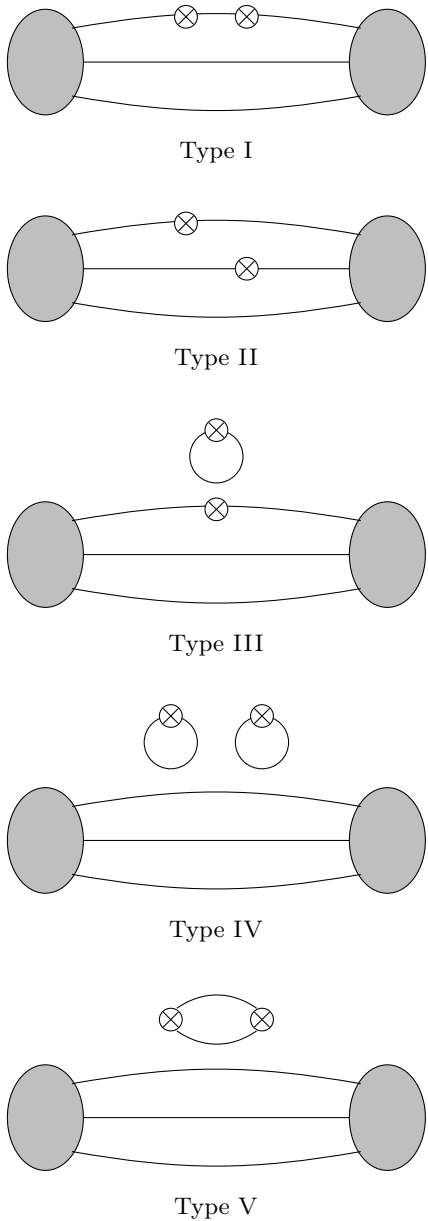
\begin{figure}[t]  % 't' forces the figure to the top of the page/column
    \centering

    \begin{tikzpicture}
    % Left ellipse
    \draw[fill=gray!50] (0,0) ellipse[x radius=0.5cm, y radius=0.7cm];

    % Right ellipse
    \draw[fill=gray!50] (4.5,0) ellipse[x radius=0.5cm, y radius=0.7cm];

    % Connecting lines
    \draw[thin] (0.35,0.45) to[out=10, in=170] (4.15,0.45); % Curved top line
    \draw[thin] (0.5,0) -- (4,0);     % Middle line
    \draw[thin] (0.35,-0.45) to[out=350, in=190] (4.15,-0.45); % Bottom line

    % Circles
    \draw[fill=white] (1.85,0.6) circle[radius=0.15cm]; % Left
    \draw[fill=white] (2.65,0.6) circle[radius=0.15cm]; % Right

    % Crosses
    \draw[thin] (1.85-0.1,0.6-0.1) -- (1.85+0.1,0.6+0.1); % BL-TR
    \draw[thin] (1.85-0.1,0.6+0.1) -- (1.85+0.1,0.6-0.1); % TL-BR

    \draw[thin] (2.65-0.1,0.6-0.1) -- (2.65+0.1,0.6+0.1); % BL-TR
    \draw[thin] (2.65-0.1,0.6+0.1) -- (2.65+0.1,0.6-0.1); % TL-BR

    % Text
    \node at (2.25,-1.2) {Type \textsc{I}};
\end{tikzpicture}

%________________________________________________________________________

\vspace{0.5cm} % Vertical space between diagrams

\begin{tikzpicture}
    % Left ellipse
    \draw[fill=gray!50] (0,0) ellipse[x radius=0.5cm, y radius=0.7cm];

    % Right ellipse
    \draw[fill=gray!50] (4.5,0) ellipse[x radius=0.5cm, y radius=0.7cm];

    % Connecting lines
    \draw[thin] (0.35,0.45) to[out=10, in=170] (4.15,0.45); % Curved top line
    \draw[thin] (0.5,0) -- (4,0);     % Middle line
    \draw[thin] (0.35,-0.45) to[out=350, in=190] (4.15,-0.45); % Bottom line

    % Circles
    \draw[fill=white] (1.85,0.6) circle[radius=0.15cm]; % Left
    \draw[fill=white] (2.65,0) circle[radius=0.15cm]; % Right

    % Crosses
    \draw[thin] (1.85-0.1,0.6-0.1) -- (1.85+0.1,0.6+0.1); % BL-TR
    \draw[thin] (1.85-0.1,0.6+0.1) -- (1.85+0.1,0.6-0.1); % TL-BR

    \draw[thin] (2.65-0.1,0-0.1) -- (2.65+0.1,0+0.1); % BL-TR
    \draw[thin] (2.65-0.1,0+0.1) -- (2.65+0.1,0-0.1); % TL-BR

    % Text
    \node at (2.25,-1.2) {Type \textsc{II}};
\end{tikzpicture}

%________________________________________________________________________

\vspace{0.5cm} % Vertical space between diagrams

\begin{tikzpicture}
    % Hollow Circle
    \draw[thin] (2.25, 1.3) circle[radius=0.35cm]; % Hollow circle at position (2.25, 1.5)

    % Left ellipse
    \draw[fill=gray!50] (0,0) ellipse[x radius=0.5cm, y radius=0.7cm];

    % Right ellipse
    \draw[fill=gray!50] (4.5,0) ellipse[x radius=0.5cm, y radius=0.7cm];

    % Connecting lines
    \draw[thin] (0.35,0.45) to[out=10, in=170] (4.15,0.45); % Curved top line
    \draw[thin] (0.5,0) -- (4,0);     % Middle line
    \draw[thin] (0.35,-0.45) to[out=350, in=190] (4.15,-0.45); % Bottom line

    % Circles
    \draw[fill=white] (2.25,0.6) circle[radius=0.15cm]; % Centre
    \draw[fill=white] (2.25,1.65) circle[radius=0.15cm]; % Top

    % Crosses
    \draw[thin] (2.25-0.1,0.6-0.1) -- (2.25+0.1,0.6+0.1); % BL-TR
    \draw[thin] (2.25-0.1,0.6+0.1) -- (2.25+0.1,0.6-0.1); % TL-BR

    \draw[thin] (2.25-0.1,1.65-0.1) -- (2.25+0.1,1.65+0.1); % BL-TR
    \draw[thin] (2.25-0.1,1.65+0.1) -- (2.25+0.1,1.65-0.1); % TL-BR

    % Text
    \node at (2.25,-1.2) {Type \textsc{III}};
\end{tikzpicture}

%________________________________________________________________________

\vspace{0.5cm} % Vertical space between diagrams

\begin{tikzpicture}
    % Hollow Circle
    \draw[thin] (1.65, 1.3) circle[radius=0.35cm]; % L circle
    \draw[thin] (2.85, 1.3) circle[radius=0.35cm]; % R circle

    % Left ellipse
    \draw[fill=gray!50] (0,0) ellipse[x radius=0.5cm, y radius=0.7cm];

    % Right ellipse
    \draw[fill=gray!50] (4.5,0) ellipse[x radius=0.5cm, y radius=0.7cm];

    % Connecting lines
    \draw[thin] (0.35,0.45) to[out=10, in=170] (4.15,0.45); % Curved top line
    \draw[thin] (0.5,0) -- (4,0);     % Middle line
    \draw[thin] (0.35,-0.45) to[out=350, in=190] (4.15,-0.45); % Bottom line

    % Circles
    \draw[fill=white] (1.65,1.65) circle[radius=0.15cm]; % Centre
    \draw[fill=white] (2.85,1.65) circle[radius=0.15cm]; % Top

    % Crosses
    \draw[thin] (1.65-0.1,1.65-0.1) -- (1.65+0.1,1.65+0.1); % BL-TR
    \draw[thin] (1.65-0.1,1.65+0.1) -- (1.65+0.1,1.65-0.1); % TL-BR

    \draw[thin] (2.85-0.1,1.65-0.1) -- (2.85+0.1,1.65+0.1); % BL-TR
    \draw[thin] (2.85-0.1,1.65+0.1) -- (2.85+0.1,1.65-0.1); % TL-BR

    % Text
    \node at (2.25,-1.2) {Type \textsc{IV}};
\end{tikzpicture}

%________________________________________________________________________

\vspace{0.5cm} % Vertical space between diagrams

\begin{tikzpicture}
    % Left ellipse
    \draw[fill=gray!50] (0,0) ellipse[x radius=0.5cm, y radius=0.7cm];

    % Right ellipse
    \draw[fill=gray!50] (4.5,0) ellipse[x radius=0.5cm, y radius=0.7cm];

    % Connecting lines
    \draw[thin] (0.35,0.45) to[out=10, in=170] (4.15,0.45); % Curved top line
    \draw[thin] (0.5,0) -- (4,0);     % Middle line
    \draw[thin] (0.35,-0.45) to[out=350, in=190] (4.15,-0.45); % Bottom line

    % Circles
    \draw[fill=white] (1.65,1.3) circle[radius=0.15cm]; % Centre
    \draw[fill=white] (2.85,1.3) circle[radius=0.15cm]; % Top

    % Additional lines
    \draw[thin] (1.73,1.45) to[out=35, in=145] (2.78,1.45);
    \draw[thin] (1.73,1.15) to[out=325, in=215] (2.78,1.15);
    
    % Crosses
    \draw[thin] (1.65-0.1,1.3-0.1) -- (1.65+0.1,1.3+0.1); % BL-TR
    \draw[thin] (1.65-0.1,1.3+0.1) -- (1.65+0.1,1.3-0.1); % TL-BR

    \draw[thin] (2.85-0.1,1.3-0.1) -- (2.85+0.1,1.3+0.1); % BL-TR
    \draw[thin] (2.85-0.1,1.3+0.1) -- (2.85+0.1,1.3-0.1); % TL-BR

    % Text
    \node at (2.25,-1.2) {Type \textsc{V}};
\end{tikzpicture}

    \caption{Lattice diagrams for all possible types of current insertion. Only Type \textsc{I} and Type \textsc{II} are considered here.}  % Caption for the diagram
    \label{fig:lattice_diags}  % Label for referencing the figure
\end{figure}

%_______________________________________________________________________
%_______________________________________________________________________

In terms of the diagrams in Fig.~\ref{fig:lattice_diags}, the different flavour combinations receive the contributions,\begin{itemize}
    \item $uu$: Type I -- \textsc{V},
    \item $dd$: Type I, \textsc{III} -- \textsc{V}
    \item $ud$: Type II -- \textsc{IV}.
\end{itemize}
This distinction is particularly relevant to our discussion of lattice artefacts. In the present work, we only consider modifications to the action on the quark propagators, but not the shift in the fermion determinant which will affect the gauge field sampling. Our calculation therefore only considers diagrams of Type I and Type II in Fig.~\ref{fig:lattice_diags}. In addition, Type \textsc{III} and Type \textsc{IV} contributions can be ignored in the present calculation, as they vanish at the SU(3) symmetric point. Since heavier quark contributions only appear in the final three types of Fig.~\ref{fig:lattice_diags}, Eq.~\eqref{proton_sub} need only include the up and down quark contributions. 

\subsection*{Lattice determination}

To determine the flavour-diagonal and flavour-mixed Compton amplitude subtraction functions, we calculate the following ratios of nucleon correlators,
\begin{align}  
        \mathcal{R}^{uu}_\lambda &= \frac{\mathcal{G}(\lambda,0)\mathcal{G}(-\lambda,0)}{[\mathcal{G}(0,0)]^2} \xrightarrow{t \gg a} A^{uu}e^{-2\Delta E_\lambda^{uu}t} \nonumber \\
        \mathcal{R}^{dd}_\lambda &= \frac{\mathcal{G}(0,\lambda)\mathcal{G}(0,-\lambda)}{[\mathcal{G}(0,0)]^2} \xrightarrow{t \gg a} A^{dd}e^{-2\Delta E_\lambda^{dd}t} \label{ratios}\\
        \mathcal{R}^{ud}_\lambda &= \frac{\mathcal{G}(\lambda,\lambda)\mathcal{G}(-\lambda,-\lambda)}{\mathcal{G}(\lambda,-\lambda)\mathcal{G}(-\lambda,\lambda)} \xrightarrow{t \gg a} A^{ud}e^{-4\Delta E_\lambda^{ud}t} \nonumber
\end{align}
as a function of Euclidean time $t$ for each value of the external coupling strength $\lambda$. Then the energy shifts are
\begin{align}
    \begin{split}
        \Delta E^{uu}_{\lambda}&=\lambda^2\left. \frac{\partial^2 E(\lambda,0)}{\partial \lambda^2}\right|_{\lambda={0}}+\mathcal{O}\left(\lambda^4\right),\\
        \Delta E^{dd}_{\lambda}&=\lambda^2\left. \frac{\partial^2 E(0,\lambda)}{\partial \lambda^2}\right|_{\lambda={0}}+\mathcal{O}\left(\lambda^4\right),\\
        \Delta E^{ud}_{\lambda}&=\lambda^2\left.\frac{\partial^2 E(\lambda_1,\lambda_2)}{\partial \lambda_1\partial\lambda_2 }\right |_{\lambda_1=\lambda_2=0}+\mathcal{O}\left(\lambda^4\right),
        \label{eshift}
    \end{split}
\end{align}
which from Eq.~\eqref{energy} are directly proportional to the subtraction function.

%\textcolor{blue}{Further detail if low $Q^2$ fits are performed.}

\subsection*{Subtraction Function Results}

\begin{table*}[t!]
    \caption{Lattice parameters and other relevant quantities of the gauge ensembles used in this work \cite{configs}, including $m_p$ \cite{Batelaan:2023aac}, $Z_V$ \cite{jacobthesis}, $g_S$ \cite{Rose} and $\langle x \rangle$ \cite{Dragos}.}
    \centering
\[
\begin{array}{c c c c c c c c c c c c c}
     N_L^3 \times N_T & a \, [\text{fm}] & \beta & \kappa_\ell, \kappa_s & am_p & am_0 & m_{\pi} \, [\text{MeV}] & m_\pi L & Z_V & g^u_S & g^d_S & \langle x \rangle_u & \langle x \rangle_d \\ 
    \hline \hline
    32^3 \times 64 & 0.074 & 5.50 & 0.120900 & 0.4646(43) & 0.00674(7) & 470 & 5.6 & 0.85 & 4.26(10) & 2.77(7) & - & -\\  
    48^3 \times 96 & 0.068 & 5.65 & 0.122005 & 0.3860(44) & 0.00545(15) & 410 & 6.8 & 0.86 & 4.83(23) & 3.15(14) & 0.403(1) & 0.173(5)\\ 
    48^3 \times 96 & 0.058 & 5.80 & 0.122810 & 0.3552(27) & 0.00574(5) & 430 & 6.1 & 0.88 & 4.47(28) & 2.90(17) & - & -\\  
\end{array}
\]
    \label{tab:ensemble_details}
\end{table*}

We perform our calculation of the proton subtraction function on three gauge ensembles generated by the QCDSF collaboration \cite{configs} using a non-perturbative, $\mathcal{O}(a)$-improved clover fermion action with unimproved current operators \cite{Cundy_2009}. All ensembles have $2+1$ quark flavours at the SU(3) flavour symmetric point, and have an unphysical pion mass---approximately three times the physical pion mass. In expectation of the presence of large discretisation artefacts, we utilise three ensembles with relatively fine values of the lattice spacing: $a=0.074,0.068,0.058\;\text{fm}$. See Table \ref{tab:ensemble_details} for more details. 

For each gauge ensemble, we determine the subtraction function for a set of $Q^2$ values given in Table \ref{tab:Q2_kinematics}, which are restricted by our Feynman-Hellmann procedure to $q_4=0$. For ease in isolating the subtraction function, we have also introduced the kinematic constraint $q_3=0$. The overall range of kinematics is $1 \lesssim Q^2 \lesssim 16 \;\text{GeV}^2$, which pushes below the threshold of the perturbative region, and up to the region where we expect the only significant contributions to be from the OPE, Eq.~\eqref{hillpaz_s1}. To determine the subtraction function we extract the energy shifts for each 3-momentum $\mathbf{q}$ of the external current, following the methodology given in Section \ref{sec:feynhell} for two values of the external coupling $\lambda \in [0.0125, 0.0250]$ as has been done in the same way in previous Feynman-Hellmann calculations of the Compton amplitude \cite{fwdpaper, fwdpaperscaling}. A single plateau fit is performed to each ratio given in Eq.~\eqref{ratios} with the plateau regions chosen following a covariance-matrix based $\chi^2$ analysis. This allows us to perform simple quadratic fits in $\lambda$ to determine the subtraction functions following Eqs.~\eqref{energy} and~\eqref{eshift}.
% USED TO BE 0.125, 0.250 FOR LAMBDA, JAMES ADDED 0 - CHECK

\begin{table}
\centering
\caption{Momentum transfers of the external current.}
	\begin{tabular}{ccc}
 \begin{tabular}[t]{cc}
 \multicolumn{2}{c}{$\kappa=0.120900$} \\
  \multicolumn{2}{c}{$\beta=5.50$} \\
		\hline\hline
			 $\mathbf{q} $ & $Q^2$ \\
    			 $[2\pi/L] $ & $[\mathrm{GeV}^2]$ \\
			\hline\hline
        $(2,0,0)$ & $1.10$  \\  
         $(2,1,0)$ & $1.37$  \\ 
          $(3,1,0)$ & $2.74$  \\ 
           $(3,2,0)$ & $3.56$  \\ 
            $(4,0,0)$ & $4.39$  \\ 
             $(4,1,0)$ & $4.66$  \\ 
              $(4,2,0)$ & $5.48$  \\ 
               $(4,3,0)$ & $6.85$  \\ 
                $(5,1,0)$ & $7.13$  \\ 
                 $(6,3,0)$ & $12.3$  \\ 
                  $(7,3,0)$ & $15.9$  \\ 
			\hline\hline
		\end{tabular}
		\quad 
		 \begin{tabular}[t]{cc} \multicolumn{2}{c}{$\kappa=0.122005$} \\
  \multicolumn{2}{c}{$\beta=5.65$} \\
		\hline\hline
			 $\mathbf{q} $ & $Q^2$ \\
    			 $[2\pi/L] $ & $[\mathrm{GeV}^2]$ \\
		\hline\hline
        $(3,1,0)$ & $1.44$  \\ 
        $(3,2,0)$ & $1.88$  \\ 
        $(4,2,0)$ & $2.89$  \\ 
        $(5,3,0)$ & $4.91$  \\ 
        $(7,1,0)$ & $7.21$  \\ 
        $(7,4,0)$ & $9.38$  \\ 
			\hline\hline
		\end{tabular}
		\quad 
		 \begin{tabular}[t]{cc}
 \multicolumn{2}{c}{$\kappa=0.122810$} \\
  \multicolumn{2}{c}{$\beta=5.80$} \\
		\hline\hline
			 $\mathbf{q} $ & $Q^2$ \\
    			 $[2\pi/L] $ & $[\mathrm{GeV}^2]$ \\
			\hline\hline
        $(4,3,0)$ & $4.96$  \\
        $(5,3,0)$ & $6.74$  \\
        $(7,1,0)$ & $9.92$  \\
			\hline\hline
		\end{tabular}
	\end{tabular}
    \label{tab:Q2_kinematics}
\end{table}

\begin{figure}
    \centering
    \includegraphics[width=\linewidth]{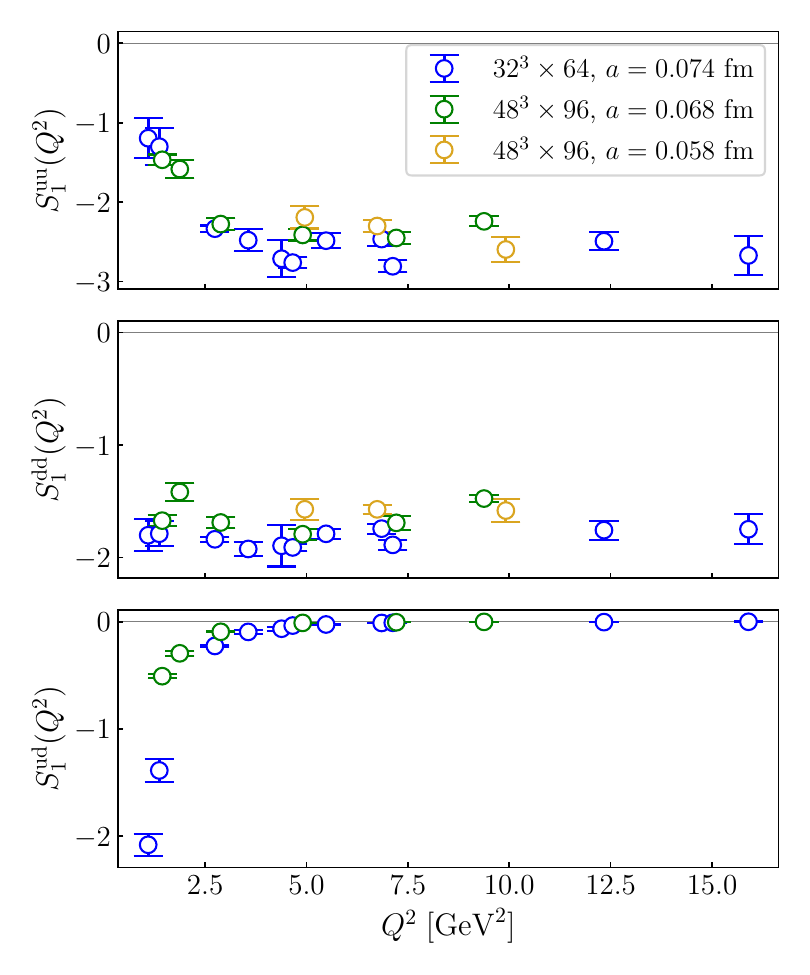}
    \caption{The lattice subtraction function as determined from Feynman-Hellmann. Results are labelled according to which quark flavour the currents couple to: $u$ for both (top), $d$ for both (middle), and one on each (bottom).}
    \label{fig:unimproved}
\end{figure}

In Fig.~\ref{fig:unimproved}, we present our results for the $uu$, $dd$ and $ud$ components of the lattice subtraction function. We can see in both the $uu$ and $dd$ results that, instead of the expected asymptotic behaviour $S_1(Q^2)\sim Q^{-2}$, the results appear to trend towards a large non-zero constant: $S_1^{uu}(Q^2)\approx -2.5$ and $S_1^{dd}(Q^2)\approx -1.5$. In contrast, the $ud$ result trends asymptotically to zero very quickly for $Q^2 \gtrsim 4$. These results are discussed in the following sections.

Although the lattice spacing varies by more than $20\%$ across the three sets of gauge configurations ($a=0.074,0.068,0.058\;\text{fm}$), the trend of our values becoming smaller as the lattice spacing becomes finer is rather subtle, and difficult to discern. Such a result naturally suggests the presence of discretisation artefacts which converge somewhat slowly.

\section{Discretisation artefacts} 
\label{sec:lope}

As with all lattice calculations, it is necessary to take into account, and correct for, discretisation artefacts present in the results. The subtraction function, more so than other quantities, suffers from such artefacts to a large degree. This is analogous to short-distance artefacts studied in the case of the hadronic vacuum polarisation \cite{HVP_shortdistance, HVP_lattartefacts2}.

To correct these artefacts, we perform a lattice operator product expansion (LOPE) of the Compton amplitude for Wilson fermions. This calculation is essentially a tree-level OPE, where the intermediate quark propagators are taken to be Wilson fermions. This section is inspired by, and builds upon, a previous LOPE by some of us in 2006 of the lattice Compton amplitude for the conserved current \cite{analytic_wilson_coefficients}. In contrast to this prior result, our current LOPE is specifically for the local current $V_{\mu} = Z_V \bar{\psi}\gamma_{\mu}\psi$ in order to match the Feynman-Hellmann calculation. Furthermore, we improve on the 2006 result by retaining the bare quark mass.

A succinct outline of the method we use to perform this LOPE is as follows (see Appendix \ref{appendix:lope} for further details): first, we expand and discretise the forward Compton amplitude ($T_{\mu\nu} \rightarrow T^L_{\mu\nu}$), then insert the Wilson form of the fermion propagators. Next, we begin the OPE by performing a Taylor expansion in small $\sin(ap_\mu)$ (the lattice equivalent of $ap_\mu$ in the continuum) and substitute this expansion into the Compton amplitude ($T_{\mu\nu}^L \rightarrow T_{\mu\nu}^{\text{LOPE}})$. Unlike the standard OPE, which uses a hard scale $Q^2 \rightarrow \infty$, we use a soft scale with $\sin(ap_\mu)$ much smaller than the momentum transfer $q$. After much simplification, we associate the $\sin(ap_\mu)$ terms with the covariant derivative before concluding by writing the relevant terms as symmetric and traceless operators. 

In our full calculation in Appendix \ref{appendix:lope}, we restrict ourselves to terms up to first order in the covariant derivative. Then, to isolate the contribution to the subtraction function, we select the $\mu=\nu=3$ component with the constraint $q_3=0$ and $p_3=0$, which follows from the lattice sink momentum $\mathbf{p} = \mathbf{0}$. Recalling the form of Eq.~\eqref{tensdecomp}, this unique selection (with $\omega = 0$, since $q^4=0$ and $\mathbf{p} = \mathbf{0}$) ensures that $T^{\text{LOPE}}_{33}$ is, in fact, the LOPE of the lattice Compton subtraction function. After removing all terms present in the continuum, we are finally left with a `correction term', $\Delta S_1$, which quantifies---to leading order---the discretisation artefact present in our lattice calculation of the subtraction function. That is, $\Delta S_1 = S_1^{\text{full}} - S_1^{\text{cont.}}$, where $S_1^{\text{full}}$ is the leading order LOPE result, and $S_1^{\text{cont.}}$ is the leading order continuum component of the LOPE, leaving only the discretisation contribution. This result is as follows (see Eq.~\eqref{A_final}):
\begin{equation}\label{DelS}
     \Delta S_1 = C_W(q,m_0) Z_V^2 \langle p | \bar{\psi}\psi | p \rangle,
\end{equation}
where $C_W(q,m_0)$ is the Wilson coefficient of the form 
\begin{equation}\label{CW}
    C_W(q,m_0) = \frac{2r\sum_{\rho}\left [ 1 - \cos(aq_{\rho})\right] }{\sum_{\rho}\sin^2(aq_{\rho})+  a^2M^2(q,m_0)}.
\end{equation}
with $M(k,m_0) = m_0 + \frac{r}{a}\sum_\tau \big[ 1 - \cos(ak_\tau) \big]$ and $r$ denoting the scalar Wilson coefficient. This scalar coefficient is typically set to $r=1$ in lattice simulations, but we keep it explicit in our LOPE expressions in order to keep track of $\mathcal{O}(a)$ terms arising from the Wilson term in the fermion action. Unlike standard Wilson coefficients, we also keep the bare quark mass explicit. 

The correction term in Eq.~\eqref{DelS} is comparable to the corresponding expression in Ref.~\cite{Capitani_2001_}. Under the same kinematic conditions, the 2006 result \cite{analytic_wilson_coefficients} (ignoring the seagull term) is in agreement with Eq~\eqref{DelS} in the limit $m_0 \rightarrow 0$, as anticipated. This expression in Eq.~\eqref{DelS} is also purely unphysical, as Taylor expanding in $a$ yields terms of order $\mathcal{O}(a)$, which vanish in the continuum. Furthermore, the expression in Eq.~\eqref{DelS} is specifically a Wilson discretisation artefact, $\Delta S_1 \, \propto \, r$, and thus there is an additional necessity for the term to vanish in the continuum \cite{private_comm}. We intend to further investigate the correction in light of clover fermions \cite{Capitani_2001_}, and for Ginsparg-Wilson fermions, for which $\mathcal{O}(a)$ contributions have already been determined \cite{GW_fermions_Oa}. The matrix element is given by $\<p|\bar{\psi}\psi|p\> = 2am_p g_S^{\text{bare}}$, where $g_S^{\text{bare}}$ is the unrenormalised scalar charge. Since $am_p\approx 0.4$ in our calculations, we note that these artefacts will be relatively large, even at somewhat fine lattice spacings.

Taking our lattice calculations of the subtraction function, $S_1^L(Q^2)$, we can now remove the leading order discretisation artefact, $\Delta S_1$, to produced an improved result
\begin{equation}\label{Corr}
    S^{\text{imp}}_1(Q^2) = S^{L}_1(Q^2) + \Delta S_1.
\end{equation}

Since the $\mathcal{O}(a)$ term of our LOPE arises from the quark propagator connecting the two current insertions (Type I in Fig.~\ref{fig:lattice_diags}), this correction only affects the $uu$ and $dd$ flavour combinations, but not the $ud$ combination. This quantitatively explains the behaviour of the results in Fig.~\ref{fig:unimproved}, which appears reasonable in the $ud$ case, but has completely anomalous asymptotes in the $uu$ and $dd$ plots.

\begin{figure}
    \centering
\includegraphics[width=\linewidth]{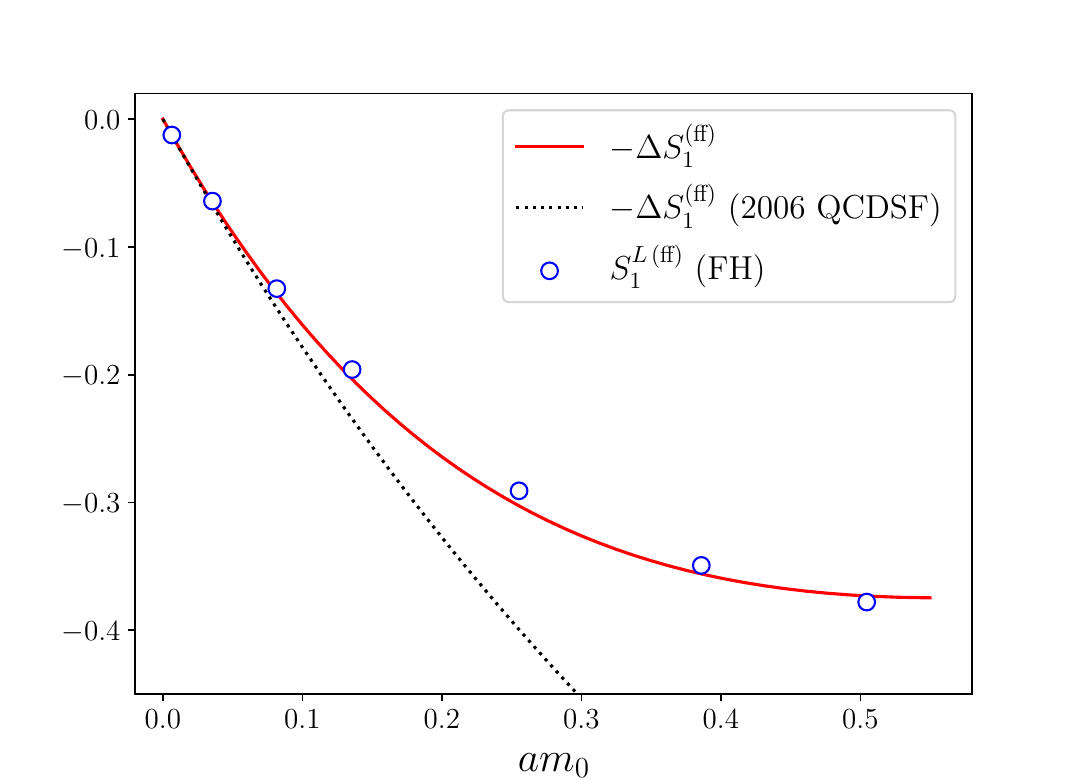}
    \caption{Comparison of our 2006 LOPE \cite{analytic_wilson_coefficients}, our lattice OPE, $\Delta S^{(\mathrm{ff})}_1$, in Eq.~\eqref{Sff}, and the Feynman-Hellmann results, ${S}_1^{L \, (\mathrm{ff})}$. Results are for free fermions and ${\mathbf{q}}=\frac{2\pi}{L}(5,1,0)$.}
\label{fig:lope_v_FH_ff.pdf}
\end{figure}

\begin{figure}
    \centering
\includegraphics[width=\linewidth]{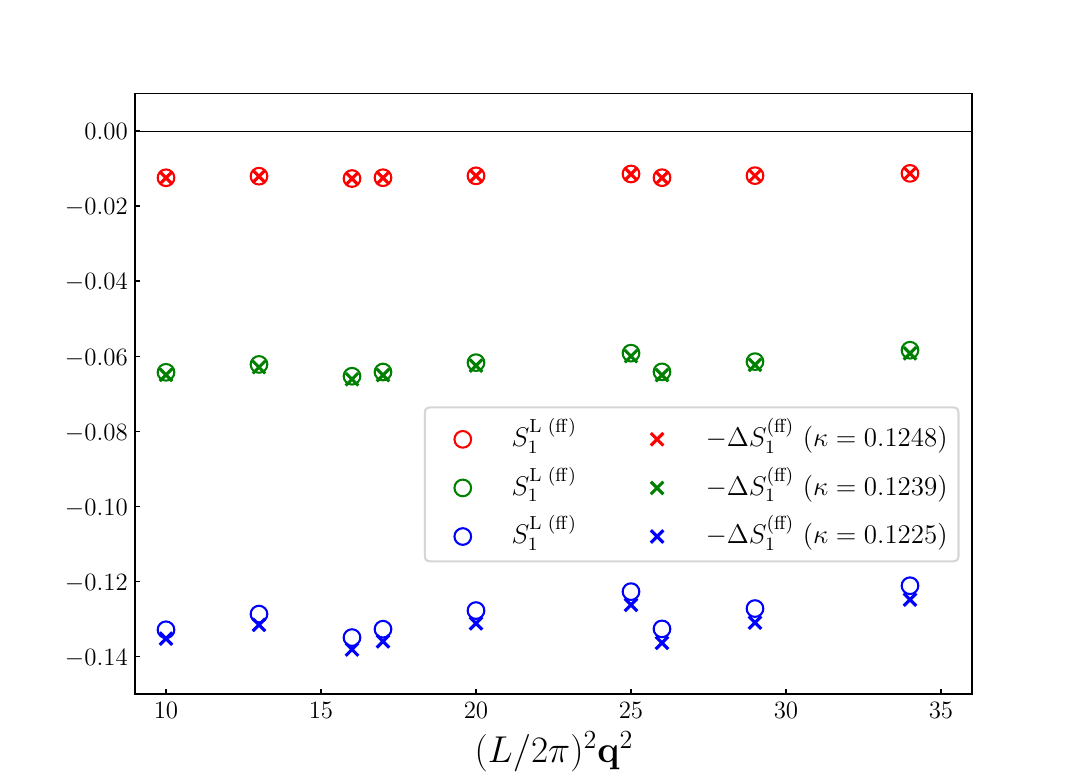}
    \caption{Comparison of the LOPE corection, $\Delta S_1^{(\text{ff})}$, and the Feynman-Hellmann (FH) results, for a free fermion over a range of ${\mathbf{q}}$ values.}
\label{fig:lope_v_FH_ff_qdep.pdf}
\end{figure}

\subsection*{Free fermion} 

To investigate the effectiveness of the LOPE correction $\Delta S_1$, we briefly consider the simple case of the free fermion. Applying simple QED, we can derive an analytic form of the forward Compton amplitude $T_{\mu\nu} $ which, under the conditions $\mu=\nu=3$ and $p_3=q_3=0$, reduces to $T_{33} = 2\omega^2/(1-\omega^2)$. At the subtraction point, $\omega=0$, the amplitude for a free fermion (ff) therefore vanishes. Setting $S_1^{\text{imp}} = 0$, Eq.~\eqref{Corr} then reduces to:
\begin{equation}\label{LOPE_v_FH}
   S_1^{L \, (\mathrm{ff})}(Q^2) =  -\Delta S^{(\mathrm{ff})}_1,
\end{equation}
where $S_1^{L \, (\mathrm{ff})}(Q^2)$ is the free fermion lattice calculation of the subtraction function, which can be determined from Feynman-Hellmann using a single perturbed propagator as in Eq.~\eqref{q_prop} (i.e.~we do not form multiple Wick contractions as in a hadron), and using unit gauge fields, $U_{\mu}=\mathbb{I}$, in the fermion matrix. We perform this calculation with a lattice size $N_L^3\times N_T = 32^3\times 64$, and for values of the hopping parameter, $\kappa$, in the range $\kappa=0.0956-0.1248$, which correspond to dimensionless bare masses in the range $am=0.0064 - 0.5$. The results of this calculation are shown as the hollow data points in Fig.~\ref{fig:lope_v_FH_ff.pdf}.
As for $\Delta S^{(\mathrm{ff})}_1$, up to higher order corrections in $a$, it takes the form
\begin{equation}\label{Sff}
    \Delta S^{(\mathrm{ff})}_1 = C_W(q,m_0) 2(am_0)g_S^{\text{bare}}
\end{equation}
where $m_0$ is the free fermion mass and we have taken $g_S^{\text{bare}}= (1+am_0)^{-1}$ and $r=1$. 
%which can be determined from the three-point calculation of $\bar{\psi}\psi$ with free Wilson fermions.

%UTKU: Naively one looks at these ranges respectively, i.e. $\kappa = 0.0956$ corresponds to $0.0064$. I think the lattice community would understand the inverse proportionality but non-lattice people would not. How about sticking with $am$ and writing something like $am=0.0064 (\kappa = 0.1248) - 1.2 (\kappa=0.0956)$ and using $am$ in figure captions? Or maybe just give $am$ because $\kappa$ feels a bit irrelevant.}

\begin{figure}[h!]
    \centering
    \includegraphics[width=\linewidth]{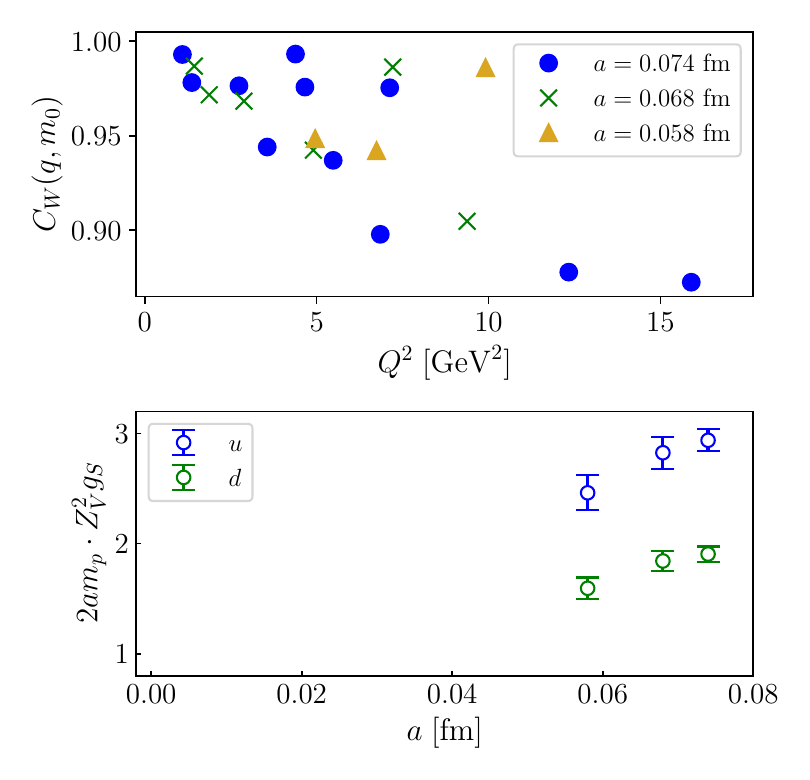}
    \caption{The two components of the leading lattice artefact to our calculation as in Eq.~\eqref{DelS}. The upper panel shows $C_W(q,m_0)$ for each kinematic point reported in Table \ref{tab:Q2_kinematics}, while the lower panel depicts $Z_V^2 \langle p | \bar{\psi}\psi | p \rangle$. The total lattice artefact is given by $\Delta S_1 = C_W(q,m_0) 2(am_p) Z_V^2 g_S$.}
    \label{fig:matelem_v_a}
\end{figure}

In Fig.~\ref{fig:lope_v_FH_ff.pdf} we plot the FH results, the discretisation artefact in Eq.~\eqref{Sff} (solid curve) and our 2006 result (broken curve). We observe that, as a function of $am_0$, the LOPE agrees well with the numerical Feynman-Hellmann results up to $am_0\approx 0.5$. Comparing this LOPE with our 2006 LOPE result (which did not include the bare mass contribution) also suggests an improvement in the correction. Similarly, in Fig.~\ref{fig:lope_v_FH_ff_qdep.pdf} we see a strong agreement between $S_1^{L \, (\mathrm{ff})}(Q^2)$ and $ \Delta S^{(\mathrm{ff})}_1$ as a function of $\mathbf{q}^2$. For the smallest $am_0$ ($\kappa=0.1248$) in particular, the agreement is excellent, and also holds to large $\mathbf{q}^2$. It's also interesting to note that, despite minor fluctuations in $\Delta S^{(\mathrm{ff})}_1$, the discretisation artefacts for the greater $am_0$ results are largely $\mathbf{q}^2$-independent.

For the free fermion we expect a strong agreement between the LOPE and numerical Feynman-Hellmann results, as $S_1^{L \, (\mathrm{ff})}(Q^2)$ vanishes in the continuum. Therefore, the agreement of these results suggest that the LOPE is a good parameterisation of our dominant discretisation artefacts, both in terms of the larger $\mathcal{O}(am_0)$ artefacts as well as the smaller $\mathcal{O}(a\mathbf{q})$ effects.

\section{Proton subtraction function} 
\label{sec:proton}

We present the application of the LOPE expressions to the $uu$ and $dd$ subtraction function results obtained in Section \ref{sec:feynhell}, and conclude by constructing the proton subtraction function. 

\begin{figure}[h!]
    \centering
    \includegraphics[width=\linewidth]{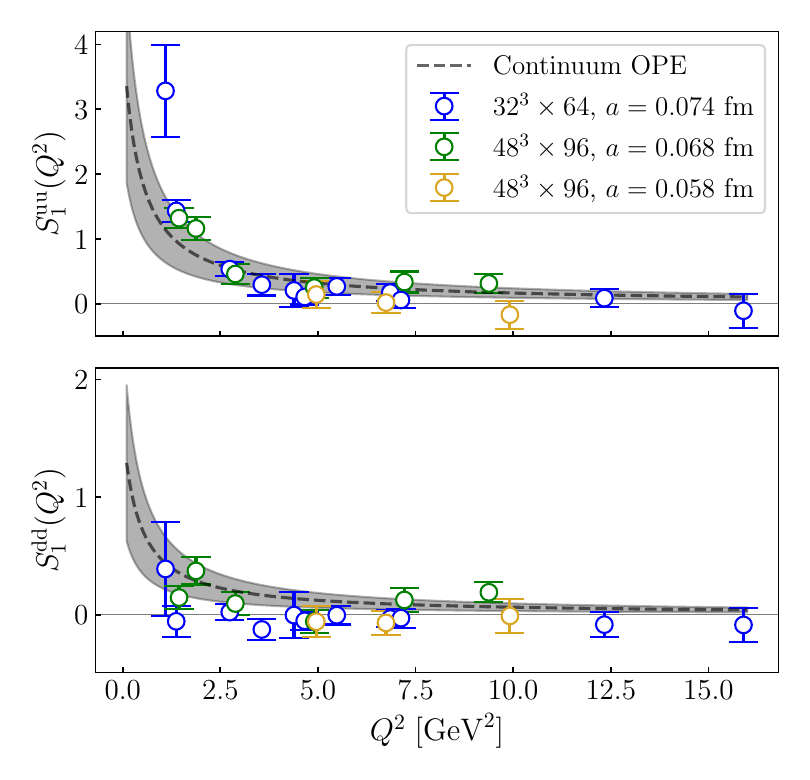}
    \caption{The subtraction function as determined from Feynman-Hellmann with lattice artefacts removed according to Eq.~\eqref{Corr}. The continuum OPE curve is also plotted, calculated directly from Eq.~\eqref{hillpaz_s1} \cite{Hill_2017} using inputs from the $a=0.074\;\text{fm}$ configurations listed in Table ~\ref{tab:ensemble_details}. As the scalar charge is flavour dependent, we present the results for both quark flavours ($u$ and $d$).}
    \label{fig:improved_uu_dd}
\end{figure}
% UTKU: Would be good to drop the Hill \& Paz reference in the figure legend. Same for the next figure.

\subsection{Evaluation of the discretisation artefact}

In Section \ref{sec:lope}, we derived a form for the leading discretisation artefact to our subtraction function, Eq.~\eqref{DelS}, which for the proton takes the form $\Delta S_1 = C_W(q,m_0) 2(am_p) Z_V^2 g^{\text{bare}}_S$. The Wilson coefficient is determined by inserting the dimensionless $a\mathbf{q}$ vectors and the bare quark masses into Eq.~\eqref{CW}. For our lattices the bare quark masses are extremely small: $am_0 = 0.00674, 0.00545, 0.00574$, for the $\beta=5.5, 5.65, 5.80$ ensembles, respectively. This makes the effects of the quark mass on $C_W(q,m_0)$ quite small. For the matrix element $Z_V^2\<p|\bar{\psi}\psi|p\>$ we determine the nucleon mass and vector renormalisation factor for each gauge ensemble separately. The bare scalar charge, $g_S^{\text{bare}}$, has been determined previously from a first-order Feynman-Hellmann calculation again for each of the ensembles \cite{Rose}. The lattice parameters are shown in Tab.~\ref{tab:ensemble_details}

In the top panel of Fig.~\ref{fig:matelem_v_a}, we plot the Wilson coefficient, $C_W(q,m_0)$, against the $Q^2$ values for all three gauge ensembles. We see in Fig.~\ref{fig:matelem_v_a} that the Wilson coefficient varies by as much as $12\%$ among the different $\mathbf{q}$ vectors, but smaller `jitters' of $\sim5\%$ are more typical.

In the bottom panel of Fig.~\ref{fig:matelem_v_a}, we plot the matrix element, $Z_V^2\<p|\bar{\psi}\psi|p\>$, against the lattice spacing $a$. We see that there is a linear dependence of the matrix element on the lattice spacing $a$. Thus we anticipate a slightly larger correction to the lattice results at a larger lattice spacing, as is suggested by Fig.~\ref{fig:unimproved}. Moreover, the `jitters' we see in the Wilson coefficient largely explain the small fluctuations with $Q^2$ in our lattice subtraction function.

\begin{figure}
    \centering
    \includegraphics[width=\linewidth]{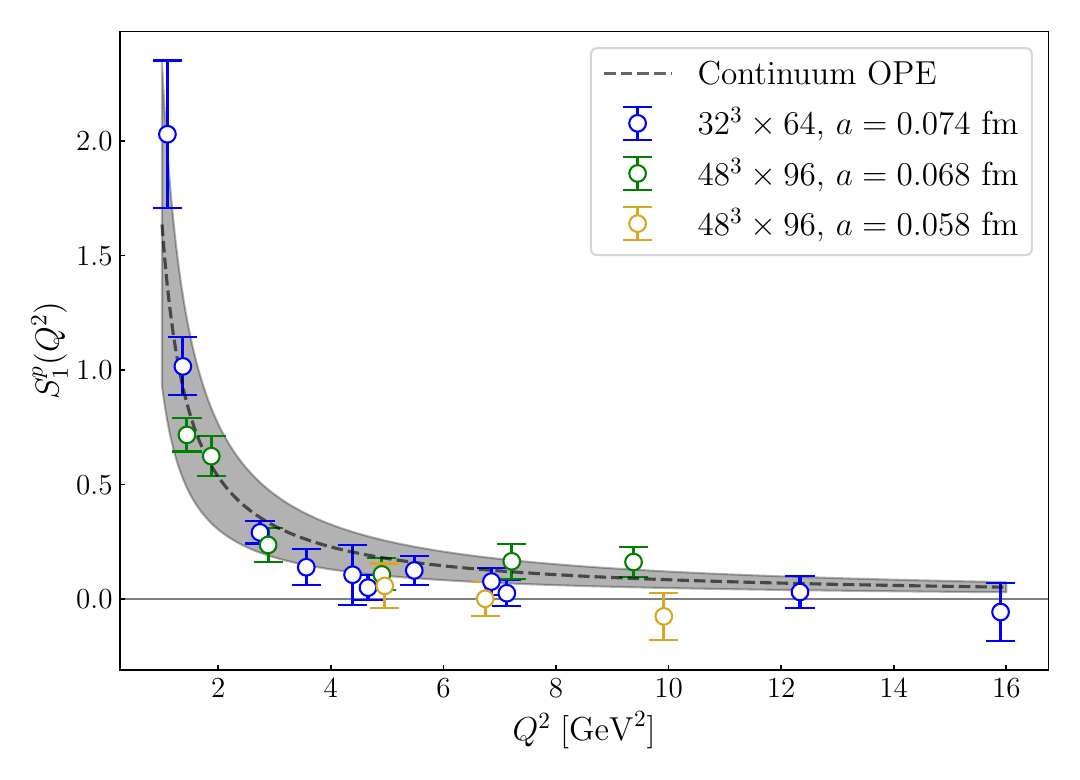}
    \caption{The proton subtraction function as determined from Feynman-Hellmann with lattice artefacts removed according to Eq.~\eqref{Corr} and models included \cite{Hill_2017}.}
    \label{fig:improved_proton}
\end{figure}

\subsection{Improved subtraction function}

Finally, we determine the improved subtraction function by removing the $\mathcal{O}(a)$ discretisation artefacts calculated in Section \ref{sec:lope} using Eq.~\eqref{Corr}. Combining the direct lattice subtraction function, Fig.~\ref{fig:unimproved}, with the leading discretisation artefacts, Fig.~\ref{fig:matelem_v_a}, we present our final results for the up ($uu$) and down ($dd$) contributions in Fig.~\ref{fig:improved_uu_dd}, and for the proton in Fig.~\ref{fig:improved_proton}. Included in both plots (dotted curve) is the continuum OPE calculated from Eq.~\eqref{hillpaz_s1} using $g_S$ and $\langle x \rangle$ from Table \ref{tab:ensemble_details}.The results presented in Fig.~\ref{fig:improved_proton} are listed in Table \ref{tab:tabulated_subfnc} of Appendix \ref{appendix:sfr}.

We immediately observe an improvement in the Fig.~\ref{fig:improved_uu_dd} results as compared to Fig.~\ref{fig:unimproved}, with the asymptotic $Q^{-2}$ behaviour restored for large momentum. Putting the results together to determine the subtraction function of the proton, Fig.~\ref{fig:improved_proton}, we observe a similar phenomenon. It appears as though $S_1(Q^2)$ follows rather simple $Q^{-2}$ behaviour down to $Q^2 \approx 1 \, \text{GeV}^2$. Given the large size of the leading discretisation artefacts in Fig.~\ref{fig:matelem_v_a}, removing these artefacts from Fig.~\ref{fig:unimproved} is sufficient to shift the points upwards into the positive domain. This upwards shift brings the results into remarkable agreement with the continuum OPE in Ref.~\cite{Hill_2017}. Furthermore, the three sets of results at different lattice spacings are very consistent with each other. 

Over the intermediate $Q^2$ values in Fig.~\ref{fig:improved_proton}, the proton subtraction function is rather well behaved, and a simple $Q^{-2}$ curve appears sufficient to capture the behaviour of the function. 

As a consequence of the different subtraction point used in Ref.~\cite{Xu_new}, we do not provide a direct comparison with the two results here, but we intend to pursue such a comparison in future work.

\section{Summary and Conclusions} \label{sec:summary}

By applying the Feynman-Hellmann method to the forward Compton amplitude, we have presented a determination of the subtraction function from first-principles. Furthermore, by performing a LOPE of the forward Compton amplitude, we successfully quantified and removed leading order Wilson discretisation artefacts, thereby placing the raw lattice results in agreement with predictions. In doing so, we demonstrated that the subtraction function indeed follows an asymptotic $Q^{-2}$ behaviour both at large $Q^2$ as well as intermediate momenta. Furthermore, the result indicates a smooth transition between the low and high $Q^2$ regimes. The results presented here establish a pathway for improved lattice calculations of the subtraction function to deliver improved estimates of the proton-neutron mass difference and the proton charge radius.

The work presented here introduces a number of avenues of future research, particularly by incorporating a consideration of finite volume effects, and by repeating this work for domain wall fermions (see Ref.~\cite{Xu_new}). Given the lack of a Wilson term, one would anticipate $\mathcal{O}(a^2)$ discretisation artefacts for domain wall fermions, as opposed to the $\mathcal{O}(a)$ artefacts present here.

The precision and applicability of our calculation could be further improved by repeating these calculations for pion masses closer to the physical value, and with additional configurations of different lattice spacings. Furthermore, the precision of the approach presented here could be evaluated by extracting $\langle x \rangle$ and $g_S$ from fits to the subtraction function, and comparing to the values in Table \ref{tab:ensemble_details}.

\section*{Acknowledgements}\label{sec:acknowledgements}

The numerical configuration generation (using the BQCD lattice QCD 
program \cite{Haar:2017ubh}) and data analysis (using the CHROMA software 
library \cite{Edwards:2004sx}) was carried out on the
DiRAC Extreme Scaling Service
(Edinburgh Centre for Parallel Computing (EPCC), Edinburgh, UK), 
the DiRAC Data Intensive Service 
(Cambridge Service for Data-Driven Discovery, CSD3, Cambridge, UK),
the Gauss Centre for Supercomputing (GCS)
(John von Neumann Institute for Computing, NIC, J\"ulich, Germany) 
and resources provided by the NHR Alliance (Berlin and G\"ottingen, Germany),
the National Computer Infrastructure 
(NCI National Facility in Canberra, Australia supported by the
Australian Commonwealth Government), the Pawsey
Supercomputing Centre, which is supported by the Australian Government and the Government of Western Australia and
the Phoenix HPC service (University of Adelaide). 
AHG and TS are supported by an Australian Government Research Training Program (RTP) Scholarship. 
R.H. is supported in part by the STFC Grant No. ST/X000494/1.
P.E.L.R. is supported in part by the STFC Grant No. ST/G00062X/1.
G.S. is supported by DFG Grant No. SCHI 179/8-1.
R.D.Y., J.M.Z. and K.U.C. are supported by the ARC Grants
No. DP190100298 and No. DP220103098 and No. DP240102839. For the purpose of open access, the authors have applied a Creative Commons Attribution (CC BY) licence to any author accepted manuscript version arising from this submission.

\appendix

\onecolumngrid

\pagebreak
\section{Details of Lattice Operator Product Expansion}
\label{appendix:lope}

Applying the fundamental thesis of the Operator Product Expansion (OPE), the time-ordered product of two local operators, $\mathcal{O}_1(z)$ and $\mathcal{O}_2(0)$, can be represented by a single local operator in the limit $z \rightarrow 0$. Writing this single operator in terms of a standard basis of operators, $\{\mathcal{O}_n\}$, the OPE can be expressed as
\begin{equation}\label{OPE_Def}
    \lim_{z \rightarrow 0} \mathcal{T} \{\mathcal{O}_1(z) \mathcal{O}_2(0) \} = \sum_n C_n(z)\mathcal{O}_n(0),
\end{equation}
where the Wilson coefficients $C_n(z)$ are c-number functions. Provided the external momentum of the system is small compared to the separation scale $1/z$, the perturbative, short-distance physics is \textit{entirely} encoded within the Wilson coefficients, while the non-perturbative, long-distance physics is \textit{entirely} contained within the operators. Of course, since QCD is an asymptotically free theory, the strong coupling constant $\alpha_s$ is small at the scale of $z$ and thus the Wilson coefficients $C_n(z)$ can be computed \textit{perturbatively}. 

In practice, Eq.~\eqref{OPE_Def} is replaced with its momentum-space counterpart
\begin{equation}\label{OPE_M_Def}
    \int d^4z \, e^{i q \cdot z} \, \mathcal{T}\{ \mathcal{O}_1(z) \mathcal{O}_2(0) \} = \sum_n C_n(q)\mathcal{O}_n(0),
\end{equation}
where $C_n(q)$ are the Fourier-transformed Wilson coefficients. To ensure the $C_n(q)$ coefficients exclusively encode the perturbative, short-distance physics, $q$ must be large compared to the momentum of the external state. As for the standard basis in the continuum, $\{ \mathcal{O}_n \}$, it is conventional to select a basis where each operator is fully symmetric and traceless, thereby ensuring linear independence and preventing cross-dimensional operator mixing. 

Turning to the forward Compton amplitude, the ultimate objective of this appendix is to perform a lattice Operator Product Expansion (LOPE) for a proton. By the very nature of the OPE, the external state of the system only contributes to the matrix elements, not the operators or Wilson coefficients themselves. Thus, for the sake of clarity, we can perform the calculation with a single quark external state of momentum $p^\mu$ with charge $\mathcal{Q}=1$, and generalise to a proton at the end by adding quark flavours and associated charges. 

To begin, recall the definition of the spin-averaged forward Compton amplitude:
\begin{equation}\label{CA}
    T_{\mu\nu} = i\int d^4z \, e^{iq \cdot z} \langle p | \mathcal{T} \{ V_\mu(z) V_\nu(0) \} | p \rangle.
\end{equation}
Wick rotating to Euclidean space, Eq. \eqref{CA} becomes \cite{firstlatticeope}
\begin{equation}\label{CA_E}
    T^{\mathcal{E}}_{\mu\nu} = \int d^4 z^{\mathcal{E}} \, e^{-i(q \cdot z)^{\mathcal{E}}} \langle p | \mathcal{T} \{ V^{\mathcal{E}}_\mu(z) V^{\mathcal{E}}_\nu(0) \} | p \rangle,
\end{equation}
where $V_\mu^\mathcal{E} = \bar{\psi}\gamma_\mu^\mathcal{E}\psi$. For the sake of convenience, we will suppress the explicit Euclidean notation moving forward. At tree-level and zeroth order in the strong coupling constant $\alpha_s$, only two contraction terms contribute to Eq. \eqref{CA_E} and thus the Compton amplitude can be written as:
\begin{equation}\label{CA_E3}
    T^L_{\mu\nu} = \bar{u}(p)\gamma_\mu \tilde{S}_W(p+q)\gamma_\nu u(p) + \bar{u}(p)\gamma_\nu \tilde{S}_W(p-q)\gamma_\mu u(p).
\end{equation}
Here we have defined the dimensionless Wilson propagator
\begin{equation}\label{Wilson_fer}
    \tilde{S}_W(k) = \frac{-i\gamma_\mu \sin(ak_\mu) + aM(k,m_0)}{\sum_\mu \sin^2(ak_\mu) + a^2M^2(k,m_0)}
\end{equation}
for 
\begin{equation}\label{M_eq}
    aM(k,m_0) = am_0 + r\sum_\tau \big[ 1 - \cos(ak_\tau) \big],
\end{equation}
with $r$ denoting the Wilson parameter and $m_0$ the bare quark mass. Having now acquired a suitable form for $T_{\mu\nu}^L$, we are now in a position to commence the OPE. Firstly, consider the explicit form of the momentum-space Wilson propagators $\tilde{S}_W(p \pm q)$ from Eq. \eqref{Wilson_fer}:
\begin{equation}\label{expandedFreeProp}
    \tilde{S}_W(p \pm q) = \frac{am_0 + r\sum_\alpha \big[ 1 - \cos(ap_\alpha \pm aq_\alpha)\big] - i\sum_\alpha\gamma_\alpha \sin(ap_\alpha \pm aq_\alpha) }{\sum_\tau \sin^2(ap_\tau \pm aq_\tau) + \big[am_0 + r\sum_\tau \big( 1 - \cos(ap_\tau \pm aq_\tau) \big) \big]^2}.
\end{equation}
Expanding the trigonometric functions and isolating all $p$-dependence in $s_\mu \equiv \sin(ap_\mu)$, Eq. \eqref{expandedFreeProp} becomes
\begin{equation}\label{superExpandedFreeProp}
    \tilde{S}_W(p \pm q) = \frac{am_0 + r\sum_\alpha \big[1 \pm \sin(aq_\alpha) s_\alpha - \cos(aq_\alpha) \sqrt{1 - s_\alpha^2}\,\big] - i\sum_\alpha\gamma_\alpha \, \big[\cos(aq_\alpha) s_\alpha \pm \sin(aq_\alpha) \sqrt{1-s_\alpha^2} \, \big] }{\sum_\tau \big[ \cos(aq_\tau) s_\tau \pm \sin(aq_\tau) \sqrt{1-s_\tau^2} \, \big]^2 + \big[am_0 + r\sum_\tau \big(1 \pm \sin(aq_\tau)s_\tau - \cos(aq_\tau)\sqrt{1-s_\tau^2} \big) \big]^2},
\end{equation}
where we have used $\cos(ap_\mu) = +\sqrt{1-s_\mu^2}$, choosing the positive square root since $\cos(ap_\mu)>0$ for very small $p_\mu$. In the continuum OPE, the calculation proceeds by expanding the denominator of the propagator as a power series in $\omega = 2p\cdot q/Q^2$ in the unphysical (i.e. $\omega < 1$) region. Given the added complexity of the lattice propagator, this standard approach is unfeasible. Rather, we proceed by expanding the denominator of Eq.~\eqref{superExpandedFreeProp}, $\tilde{S}_W^{\text{dn}}(p \pm q)$, in small $p_\mu$ (hence small $s_\mu$). Doing so, we find
\begin{align}\label{DenomS_W}
    \tilde{S}_W^{\text{dn}}(p \pm q) &\equiv \frac{1}{\sum_\tau \big[ \cos(aq_\tau) s_\tau \pm \sin(aq_\tau) \sqrt{1-s_\tau^2} \, \big]^2 + \big[am_0 + r\sum_\tau \big(1 \pm \sin(aq_\tau)s_\tau - \cos(aq_\tau)\sqrt{1-s_\tau^2} \big) \big]^2} \nonumber \\
    &= \frac{1}{\Omega(q,m_0)} \mp \frac{1}{\Omega^2(q,m_0)} \sum_\tau 2\sin(aq_\tau)  \big[\cos(aq_\tau) + raM(q,m_0)  \big]s_\tau + \mathcal{O}(s^2),
\end{align}
where we define
\begin{equation}
    \Omega(q,m_0) = \sum_\rho \sin^2(aq_\rho) + a^2M^2(q,m_0).
\end{equation}
Substituting Eq.~\eqref{DenomS_W} and Eq.~\eqref{superExpandedFreeProp} back into Eq.~\eqref{CA_E3}, and further expanding the numerator terms $\sqrt{1-s_\tau^2} = 1 + \mathcal{O}(s^2 )$, we arrive at the following form:

\begin{align}\label{Latt_first_exp}
    T^{\text{LOPE}}_{\mu\nu} &= \frac{1}{\Omega(q,m_0)} \bar{u}(p) \gamma_\mu \big[ -i\sum_\alpha \cos(aq_\alpha)\gamma_\alpha s_\alpha - i\sum_\alpha \sin(aq_\alpha)\gamma_\alpha + r\sum_\alpha \sin(aq_\alpha)s_\alpha + aM(q,m_0)\big]\gamma_\nu \nonumber\\
    &\times \bigg[ 1 - \frac{2}{\Omega(q,m_0)}\sum_\tau \sin(aq_\tau)  \big[\cos(aq_\tau) + raM(q,m_0)  \big]s_\tau \bigg]u(p) \nonumber\\
    &+\frac{1}{\Omega(q,m_0)} \bar{u}(p) \gamma_\nu \big[ -i\sum_\alpha\cos(aq_\alpha)\gamma_\alpha s_\alpha + i\sum_\alpha \sin(aq_\alpha)\gamma_\alpha - r\sum_\alpha \sin(aq_\alpha)s_\alpha + aM(q,m)\big]\gamma_\mu \nonumber\\
    &\times \bigg[ 1 + \frac{2}{\Omega(q,m_0)} \sum_\tau \sin(aq_\tau)  \big[\cos(aq_\tau) + raM(q,m_0)  \big]s_\tau \bigg]u(p) + \mathcal{O}(s^2).
\end{align}
Rewriting Eq.~\eqref{Latt_first_exp} yields
\begin{align}\label{Latt_sec_exp}
    T_{\mu\nu}^{\text{LOPE}} &= \frac{1}{\Omega(q,m_0)} \bar{u}(p) \bigg[ -i\gamma_\mu \sum_\alpha [\cos(aq_\alpha)\gamma_\alpha s_\alpha + \sin(aq_\alpha)\gamma_\alpha] \gamma_\nu -i \gamma_\nu \sum_\alpha [\cos(aq_\alpha)\gamma_\alpha s_\alpha - \sin(aq_\alpha)\gamma_\alpha ] \gamma_\mu\nonumber\\
    &\;\;\;\;\;\;\;\;\;\;\;\;\;\;\;\;\;\;\;\;\;\;\;\;\;\;\; + r\sum_\alpha \sin(aq_\alpha)s_\alpha (\gamma_\mu\gamma_\nu - \gamma_\nu\gamma_\mu) + aM(q,m_0)(\gamma_\mu\gamma_\nu + \gamma_\nu\gamma_\mu) \bigg]u(p)\nonumber\\
    &+ \frac{2}{\Omega^2(q,m_0)}\sum_\tau X_\tau(q,m_0) \bar{u}(p) \bigg[ i\sum_\alpha \sin(aq_\alpha)[\gamma_\mu\gamma_\alpha \gamma_\nu + \gamma_\nu \gamma_\alpha \gamma_\mu]s_\tau\nonumber\\
    &\;\;\;\;\;\;\;\;\;\;\;\;\;\;\;\;\;\;\;\;\;\;\;\;\;\;\;\;\;\;\;\;\;\;\;\;\;\;\;\;\;\;\;\;\;\;\;\;\;\; - aM(q,m_0)(\gamma_\mu\gamma_\nu - \gamma_\nu\gamma_\mu) s_\tau \bigg] u(p) + \mathcal{O}(s^2),
\end{align}
where $X_\tau(q,m_0) = \sin(aq_\tau) \big[ \cos(aq_\tau) + raM(q,m_0) \big]$. Applying the anti-commutation relation for Euclidean gamma matrices, $\{\gamma_\mu, \gamma_\nu \} = 2\delta_{\mu\nu}\mathbb{I}$, and the identity $\gamma_\mu\gamma_\tau\gamma_\nu = \delta_{\mu\tau}\gamma_\nu + \delta_{\tau\nu}\gamma_\mu - \delta_{\mu\nu}\gamma_\tau + \epsilon_{\sigma\mu\tau\nu}\gamma_\sigma \gamma_5$, we can write Eq.~\eqref{Latt_sec_exp} as:
\begin{align}\label{Latt_third_exp}
    T^{\text{LOPE}}_{\mu\nu} &= \frac{-2i}{\Omega(q,m_0)} \bar{u}(p) \bigg[ \cos(aq_\mu)s_\mu\gamma_\nu + \cos(aq_\nu)s_\nu\gamma_\mu - \delta_{\mu\nu}\sum_\alpha \cos(aq_\alpha)s_\alpha \gamma_\alpha +\sum_\alpha \epsilon_{\sigma\mu\alpha\nu}\sin(aq_\alpha)\gamma_\sigma\gamma_5 \nonumber\\
    &\;\;\;\;\;\;\;\;\;\;\;\;\;\;\;\;\;\;\;\;\;\;\;\;\;\;\; + r\sum_\alpha \sin(aq_\alpha)s_\alpha \sigma_{\mu\nu} + iaM(q,m_0)\delta_{\mu\nu} \bigg]u(p) \nonumber\\
    &+ \frac{4i}{\Omega^2(q,m_0)} \sum_\tau X_\tau(q,m_0) \bar{u}(p) \bigg[ \sin(aq_\mu)s_\tau\gamma_\nu + \sin(aq_\nu)s_\tau\gamma_\mu - \delta_{\mu\nu} \sum_\alpha \sin(aq_\alpha)s_\tau \gamma_\alpha \nonumber\\
    &\;\;\;\;\;\;\;\;\;\;\;\;\;\;\;\;\;\;\;\;\;\;\;\;\;\;\;\;\;\;\;\;\;\;\;\;\;\;\;\;\;\;\;\;\;\;\;\;\;\; + aM(q,m_0)s_\tau \sigma_{\mu\nu} \bigg] u(p) + \mathcal{O}(s^2)
\end{align}
Thus far, we have neglected the \textit{spin-averaged} nature of the amplitude under consideration. Having sufficiently expanded the expression, we can now separate the amplitude into a symmetric and antisymmetric component: $T_{\mu\nu}^{\text{LOPE}} = T_{\{\mu\nu\}}^{\text{LOPE}} + T_{[\mu\nu]}^{\text{LOPE}}$. Then, accounting for spin-averaging, we discard the antisymmetric term, leaving
\begin{align}\label{Latt_4_exp}
    T_{\{\mu\nu\}}^{\text{LOPE}} &= \frac{-2i}{\Omega(q,m_0)} \bar{u}(p) \bigg[ \cos(aq_\mu)s_\mu\gamma_\nu + \cos(aq_\nu)s_\nu\gamma_\mu - \delta_{\mu\nu}\sum_\alpha \cos(aq_\alpha)s_\alpha \gamma_\alpha + iaM(q,m_0)\delta_{\mu\nu} \bigg]u(p) \nonumber\\
    &+ \frac{4i}{\Omega^2(q,m_0)} \sum_\tau X_\tau(q,m_0) \bar{u}(p) \bigg[ \sin(aq_\mu)s_\tau \gamma_\nu + \sin(aq_\nu)s_\tau \gamma_\mu - \delta_{\mu\nu}\sum_\alpha \sin(aq_\alpha) s_\tau \gamma_\alpha  \bigg] u(p) + \mathcal{O}(s^2).
\end{align}
In discretised Euclidean space, we associate $\sin(ap_{\mu})\leftrightarrow -ia{\overset{\leftrightarrow} D}_{\mu}$, where ${\overset{\leftrightarrow} D}_{\mu} = (\vec{D} - \cev{D})/2$, since $p_\mu$ is associated with $-i\partial_\mu$ in continuous Euclidean spacetime. The covariant derivative here is, of course, the lattice covariant derivative (i.e. finite difference operator), which reduces to the continuum covariant derivative in the limit $a\rightarrow 0$. Defining $T_{\{\mu\nu\}}^{\text{LOPE}} = \langle p | t_{\{\mu\nu \}}^{\, \text{LOPE}} | p \rangle$, Eq.~\eqref{Latt_4_exp} becomes
\begin{align}\label{Latt_5_exp}
    t_{\{\mu\nu \}}^{\, \text{LOPE}} &= \frac{2ai}{\Omega(q,m_0)} \bigg[\cos(aq_\mu)\bar{\psi}\gamma_\nu i{\overset{\leftrightarrow} D}_\mu\psi + \cos(aq_\nu)\bar{\psi}\gamma_\mu i{\overset{\leftrightarrow} D}_\nu \psi - \delta_{\mu\nu}\sum_\alpha \cos(aq_\alpha)\bar{\psi}\gamma_\alpha i{\overset{\leftrightarrow} D}_\alpha \psi  - iM(q,m_0)\delta_{\mu\nu} \bar{\psi}\psi\bigg]\nonumber\\
    &- \frac{4ai}{\Omega^2(q,m_0)} \sum_\tau X_\tau(q,m_0) \bigg[ \sin(aq_\mu)\bar{\psi}\gamma_\nu i{\overset{\leftrightarrow} D}_\tau \psi  + \sin(aq_\nu)\bar{\psi} \gamma_\mu i{\overset{\leftrightarrow} D}_\tau \psi - \delta_{\mu\nu}\sum_\alpha\sin(aq_\alpha) \bar{\psi}\gamma_\alpha i{\overset{\leftrightarrow} D}_\tau \psi  \bigg] + \mathcal{O}(D^2).
\end{align}
We are now in the position to construct \textit{symmetric} and \textit{traceless} operators of the form:
\begin{equation}\label{Operators}
    \mathcal{O}^S = \bar{\psi}\psi, \;\;\;\;\;\;\;
    \mathcal{O}_{\mu_1 ... \mu_n}^{V} = \bar{\psi} \big[ (-1)^{n-1}\gamma_{\{ \mu_1} \, i{\overset{\leftrightarrow} D}_{\mu_2} \, ... \, i{\overset{\leftrightarrow} D}_{\mu_n \}} \big] \psi - \text{traces},
\end{equation}
which we define in such a way so that, in the limit $a \rightarrow 0$, these reduce to the standard continuum operators. Unlike the continuum OPE, Eq.~\eqref{Latt_5_exp} has additional $q$-dependent trigonometric coefficients which interfere with the construction of such operators. Thus, we expand the trigonometric functions as:
\begin{align}\label{Latt_6_exp}
    t_{\{\mu\nu \}}^{\, \text{LOPE}} &= \frac{2a}{\Omega(q,m_0)} M(q,m_0) \delta_{\mu\nu} \bar{\psi}\psi \nonumber\\
    &+ \frac{2ai}{\Omega(q,m_0)} \bigg[[1-2\sin^2(aq_\mu/2)]\bar{\psi}\gamma_\nu i{\overset{\leftrightarrow} D}_\mu\psi + [1-2\sin^2(aq_\nu)]\bar{\psi}\gamma_\mu i{\overset{\leftrightarrow} D}_\nu \psi - \delta_{\mu\nu}\sum_\alpha [1 - 2\sin^2(aq_\alpha)]\bar{\psi}\gamma_\alpha i{\overset{\leftrightarrow} D}_\alpha \psi \bigg]\nonumber\\
    &- \frac{4ai}{\Omega^2(q,m_0)} \sum_\tau X_\tau(q,m_0) \bigg[ \sin(aq_\mu)\bar{\psi}\gamma_\nu i{\overset{\leftrightarrow} D}_\tau \psi  + \sin(aq_\nu)\bar{\psi} \gamma_\mu i{\overset{\leftrightarrow} D}_\tau \psi - \delta_{\mu\nu}\sum_\alpha\sin(aq_\alpha) \bar{\psi}\gamma_\alpha i{\overset{\leftrightarrow} D}_\tau \psi  \bigg] + \mathcal{O}(D^2)\nonumber\\
    &= \frac{2aM(q,m_0)}{\Omega(q,m_0)}\delta_{\mu\nu}\mathcal{O}^S + \frac{-2a^2i}{\Omega(q,m_0)}\bigg[ \mathcal{O}_{\mu\nu}^V - \frac{1}{2}\delta_{\mu\nu}\bar{\psi}i\slashed{D}\psi \bigg]\nonumber\\
    &+ \frac{-4ai}{\Omega(q,m_0)}\bigg[\sin^2(aq_\mu/2)\bar{\psi}\gamma_\nu i{\overset{\leftrightarrow} D}_\mu\psi \nonumber + \sin^2(aq_\nu)\bar{\psi}\gamma_\mu i{\overset{\leftrightarrow} D}_\nu \psi - \delta_{\mu\nu}\sum_\alpha \sin^2(aq_\alpha) \bar{\psi}\gamma_\alpha i{\overset{\leftrightarrow} D}_\alpha \psi \bigg]\nonumber\\
    &- \frac{4ai}{\Omega^2(q,m_0)} \sum_\tau X_\tau(q,m_0) \bigg[ \sin(aq_\mu)\bar{\psi}\gamma_\nu i{\overset{\leftrightarrow} D}_\tau \psi  + \sin(aq_\nu)\bar{\psi} \gamma_\mu i{\overset{\leftrightarrow} D}_\tau \psi - \delta_{\mu\nu}\sum_\alpha\sin(aq_\alpha) \bar{\psi}\gamma_\alpha i{\overset{\leftrightarrow} D}_\tau \psi  \bigg] + \mathcal{O}(D^2).
\end{align}
At leading order in $a$, $\bar{\psi}\slashed{D}\psi=m_0\mathcal{O}^S$ and thus Eq.~\eqref{Latt_6_exp} becomes:
\begin{align}\label{Latt_7_exp}
    t_{\{\mu\nu \}}^{\, \text{LOPE}} &= \frac{2a}{\Omega(q,m_0)}\delta_{\mu\nu}\bigg[ \frac{1}{2}m_0 + \frac{r}{a}\sum_\tau \big[ 1-\cos(aq_\tau) \big]\bigg]\mathcal{O}^S - \frac{2ai}{\Omega(q,m_0)}\mathcal{O}^V_{\mu\nu}\nonumber\\
    &- \frac{4ai}{\Omega(q,m_0)}\bigg[\sin^2(aq_\mu/2)\bar{\psi}\gamma_\nu i{\overset{\leftrightarrow} D}_\mu\psi \nonumber + \sin^2(aq_\nu)\bar{\psi}\gamma_\mu i{\overset{\leftrightarrow} D}_\nu \psi - \delta_{\mu\nu}\sum_\alpha \sin^2(aq_\alpha) \bar{\psi}\gamma_\alpha i{\overset{\leftrightarrow} D}_\alpha \psi \bigg]\nonumber\\
    &- \frac{4ai}{\Omega^2(q,m_0)} \sum_\tau X_\tau(q,m_0) \bigg[ \sin(aq_\mu)\bar{\psi}\gamma_\nu i{\overset{\leftrightarrow} D}_\tau \psi  + \sin(aq_\nu)\bar{\psi} \gamma_\mu i{\overset{\leftrightarrow} D}_\tau \psi - \delta_{\mu\nu}\sum_\alpha\sin(aq_\alpha) \bar{\psi}\gamma_\alpha i{\overset{\leftrightarrow} D}_\tau \psi  \bigg] + \mathcal{O}(D^2),
\end{align}
This is as far as we can reasonably progress. The terms in the second line of Eq.~\eqref{Latt_7_exp} cannot be converted into operators of the form Eq.~\eqref{Operators} as a result of interference from the trigonometric terms. However, for our purposes, we need not progress any further. Rather, we restrict our attention to only the terms at zeroth order in the covariant derivative:
\begin{align}\label{0_order}
    t_{\{\mu\nu \}}^{\, \text{LOPE}} &= \frac{2a}{\Omega(q,m_0)}\delta_{\mu\nu}\bigg[ \frac{1}{2}m_0 + \frac{r}{a}\sum_\tau \big[ 1-\cos(aq_\tau) \big]\bigg]\mathcal{O}^S + \mathcal{O}(D)\nonumber\\
    &= C_0(q,m_0) \delta_{\mu\nu} \mathcal{O}^S + \mathcal{O}(D),
\end{align}
where $C_0(q,m_0)$ is the Wilson coefficient for the zeroth order term in the Compton amplitude OPE expansion. Then the full amplitude becomes
\begin{equation}
    T_{\{\mu\nu\}}^{\text{LOPE}} = C_0(q,m_0) \delta_{\mu\nu} \langle p | \bar{\psi}\psi | p \rangle + \mathcal{O}(D) \label{EEE}
\end{equation} 
where the matrix element $\langle p | \bar{\psi}\psi | p \rangle \propto a$. There are two components to Eq.~\eqref{EEE}, a continuum component of order $\mathcal{O}(1/a)$ which, together with the matrix element, survives the continuum limit, and a lattice component which vanishes in the limit $a \rightarrow 0$. The continuum component is given by
\begin{equation}
    \lim_{a \rightarrow 0} aC_0(q,m_0) = \frac{m_0}{q^2 + m_0^2}.
\end{equation}
which simply corresponds to the first term of Eq.~\eqref{0_order}. To determine the pure `discretisation contribution' to the lattice Compton amplitude, $\Delta T_{\{\mu\nu\}}$, we can therefore simply remove this term, leaving
\begin{equation}
    \Delta T_{\{\mu\nu\}} = \frac{2r}{\Omega(q,m_0)}\delta_{\mu\nu}\sum_\tau \big[ 1-\cos(aq_\tau) \big]\langle p | \mathcal{O}^S | p \rangle + \mathcal{O}(D)
\end{equation}
which entirely vanishes in the continuum. We can apply this same result to the subtraction function by selecting the component $\mu=\nu=3$ with $\omega=0$ so that Eq.~\eqref{tensdecomp} reduces to $T_{33} = S_1(Q^2)$. Since we select the $\mu=\nu=3$ component, we also incorporate the factor of $(-i)^2$ that was neglected during the Wick rotation ($\mu=3$ component of the vector current picks up factor of $-i$ in shifting from Minkowski to Euclidean space). Finally, we introduce a factor of $(-1)$ for the sake of general clarity, and correct for this change of sign by adding the correction in Eq.~\eqref{Corr}, as opposed to subtracting it. Thus, for a quark state, we can define a `correction term' to the subtraction function,
\begin{equation}\label{A_final}
    \Delta S_1 \equiv C_W(q,m_0)\langle p | \bar{\psi}\psi | p \rangle,
\end{equation}
where
\begin{equation}
    C_W(q,m_0) \equiv \frac{2r \sum_\tau \big[ 1 - \cos(aq_\tau) \big]}{\Omega(q,m_0)}
\end{equation}
is the discretised `Wilson coefficient'. This final result can be applied to the proton by simply extending to multiple quark flavours with their associated charges, and introducing a renormalisation factor, $Z_V$, to the vector currents. It should be noted that this does not contribute additional terms in the Wick expansion of Eq.~\eqref{CA_E} from the product of two different currents because terms from these contractions are of twist $t \ge 4$, and thus do not contribute at leading order.

\pagebreak
\section{Tabulated Subtraction Function Results}
\label{appendix:sfr}

\begin{table}[H]
\centering
\caption{$\mathcal{O}(a)$-improved subtraction function results for the different configurations, as shown in Fig.~\ref{fig:improved_proton}.}
    \begin{tabular}{ccc}
    % First table
    \begin{tabular}[t]{cc}
    \multicolumn{2}{c}{$\kappa=0.120900$} \\

    $Q^2 \, [\text{GeV}^2]$ & $S_1(Q^2)$ \\ 
    \hline \hline
    1.10 & 2.00(32) \\  
    1.37 & 0.984(126)\\ 
    2.74 & 0.259(48)\\
    3.56 & 0.108(79)\\
    4.39 & 0.074(131)\\
    4.66 & 0.018(53)\\
    5.48 & 0.094(63)\\
    6.85 & 0.047(59)\\
    7.13 & -0.006(56)\\
    12.3 & 0.002(70)\\
    15.9 & -0.085(126)\\
    \hline \hline
    \end{tabular} &

    % Second table
    \begin{tabular}[t]{cc} 
    \multicolumn{2}{c}{$\kappa = 0.122005$} \\

    $Q^2 \, [\text{GeV}^2]$ & $S_1(Q^2)$ \\ 
    \hline \hline
    1.44 & 0.688(71) \\
    1.88 & 0.596(87) \\
    2.89 & 0.208(73) \\
    4.91 & 0.080(69) \\
    7.21 & 0.136(75) \\
    9.38 & 0.135(65) \\
    \hline \hline
    \end{tabular} &

    % Third table
    \begin{tabular}[t]{cc}
    \multicolumn{2}{c}{$\kappa=0.122810$} \\

    $Q^2 \, [\text{GeV}^2]$ & $S_1(Q^2)$ \\ 
    \hline \hline
    4.96 & 0.056(98)\\
    6.74 & -0.001(75)\\
    9.92 & -0.077(103)\\
    \hline \hline
    \end{tabular}
    \end{tabular}
    \label{tab:tabulated_subfnc}
\end{table}

\twocolumngrid

\bibliography{main}

\end{document}